\documentclass[11pt]{article}
\pdfoutput=1
\usepackage{setspace}
\usepackage[space]{grffile}
\usepackage{amsmath}
\usepackage{url}
\usepackage{amsthm}
\usepackage[sort,comma,authoryear]{natbib}

\usepackage{mathtools}
\usepackage{bm}
\usepackage{amssymb}
\usepackage{threeparttable}
\DeclarePairedDelimiter{\abs}{\lvert}{\rvert}
\usepackage{booktabs}
\usepackage{array}
\usepackage[english]{babel}
\usepackage{longtable}
\usepackage{subfig}
\usepackage{ragged2e}
\usepackage{rotating}
\theoremstyle{definition}

\theoremstyle{remark}

\newcommand{\tabincell}[2]{\begin{tabular}{@{}#1@{}}#2\end{tabular}}
 \usepackage[titletoc]{appendix}
\usepackage{tabularx}
\usepackage{changepage}
\usepackage{verbatim}
\usepackage[a4paper, total={6in, 8in}]{geometry}
\usepackage{setspace}
\usepackage{booktabs}
\usepackage{multirow}
\usepackage{epsf}
\usepackage{geometry}
\usepackage{graphicx}% To incorporate -eps-converted-to.pdf illustrations using PDFLaTeX, etc.
\usepackage{epstopdf}%epstopdf

\usepackage{hyperref}
\hypersetup{
	colorlinks,
	linkcolor={blue},
	citecolor={blue},
	urlcolor={blue}
}
\begin{document}

\title{Using expectile regression for classification ratemaking}
	\author{Liang Yang\footnote{School of Insurance, Southwestern University of Finance and Economics, Chengdu, China.}
		\quad Zhengxiao Li\footnote{School of Insurance and Economics, University of International Business and Economics, Beijing, China.}
		\quad Shengwang Meng\footnote{Corresponding author, Email: mengshw@ruc.edu.cn. Center for Applied Statistics, Renmin University of China, Beijing, China.}
}

	\date{}
	\maketitle

\bibliographystyle{plainnat}
	\begin{abstract}
Calculating the risk premium is a prime objective in non-life actuarial science. This paper examines an application of expectile regression
to determine risk premium rates for tariff classes.
A so-called Expectile Premium Principle that inherits the good statistical properties of the expectile regression is investigated. This new premium principle as a coherent risk measure can give more information about the shape of the entire loss distribution.
For model comparison, we consider the conventional GLMs and three quantile regression models discussed in
\citet{heras2018application} and \citet{baione2019, baione2020application} as the benchmarks.
{A simulation study is designed to
evaluate the model performance.
Based on {a real world} automobile insurance data set, our experimental result reveals the expectile regression with Expectile Premium Principle has the advantage of better differentiating the heterogeneity among tariff classes and also has a greater ability to distinguish high risks from low risks.
}
%Based on a simulation study and {a real world} automobile insurance data set, our experimental result reveals that the proposed ratemaking model outperforms the existing approaches.
%It has the advantage of better differentiating the heterogeneity among tariff classes and also has a greater ability to distinguish high risks from low risks.
%Calculating the risk premium is a prime objective in non-life actuarial science. The paper examines an application of quantile regression to risk premium calculation, the so-called Expectile Premium Principle (EPP). This new premium principle inherits the good properties of the expectile, e.g., asymptotic unbiasedness and normality.
%EPP based on an expectile measure, corresponding to individual insurance risk, can be easily proven to be a coherent risk measure, can gives more information about the shape of the entire loss distribution. For model comparison, we consider the conventional GLMs and three quantile regression models discussed in Heras et al. (2018) and Baione and Biancalana (2019, 2020) as the benchmarks. Based on a simulation study and a real world automobile insurance data set, our experimental result reveals that EPP outperforms the existing approaches. Our proposed method can better differentiate the heterogeneity among tariff classes and has a greater ability to distinguish high risks from low risks
%Moreover, it allows maintaining efficiency in the estimation of risk premiums in classification ratemaking.
%		\\
%		\textbf{JEL code:} C1
		\\	\\
	\textbf{Compiled date}: \date{\today}
	\end{abstract}
\newpage

\section{Introduction}\label{sec: introduction}

In a highly competitive insurance market, one of the most important objectives for actuaries is to determine appropriate risk premiums for tariff classes (the policyholders with similar risks are classified into the same tariff class based on various risk factors).
The risk premium usually involves a separate analysis of two parts: pure premium and risk loading. The former corresponds to the expected value of future losses and the latter is
supporting the insurer's ability to cover additional losses caused by the unfavorable deviation.
While a rich variety of literature on classification ratemaking methods has proposed how to predict the pure premium (\cite{yang2018insurance, smyth2002fitting, henckaerts2018data}, among others),
%the method for determining
the statistical estimation of the risk loading has received much less attention, especially in the presence of additional covariate information.
To estimate the risk loading correctly and at the same time allow classification by tariff features, we
develop a new statistical framework to predict risk premiums of individual policies based on an
arbitrary set of risk factors.

%In the ratemaking literature,
In order to determine the risk premium for a individual risk,
many different premium principles have been proposed in actuarial science.
The most common approach in  ratemaking for tariff classes is
to
apply the Expected Value Premium Principle (EVPP) or the
Standard Deviation Premium Principle (SDPP), in which
the pure premium is based on conventional generalized linear models (GLMs) and
the risk loading can be expressed as a certain percentage of the expectation or the standard deviation of the losses.
An alternative method is to apply a distortion function as a premium principle, from which the risk premium is calculated on the basis of the loss distribution, e.g., Wang Premium Principle \citep{wang2000class} and Value-at-risk (VaR) Premium Principle \citep{kudryavtsev2009using}.
The VaR Premium Principle is based on quantile regression model when a set of risk factors are considered.
It explains the needs of risk loading quite well, as it estimates the maximum possible losses that an individual policy may incur with a given probability $1-\tau$ during the forecasting period, in which the risk loading is expressed as the difference between the $\tau$-th quantile and $50\%$-th quantile.
%as it can allow a direct estimate of an individual policy's VaR, in which the risk loading is expressed as the difference between the $\tau$-th quantile and $50\%$-th quantile.
%consider risk premium as a whole,
%so-called Wang Premium Principle
%%The second one is to
%consider risk premium as a whole,
%especially
%including the Value-at-risk (VaR) Premium Principle that was first discussed in \cite{kudryavtsev2009using}.
%%\citep{kudryavtsev2009using} and the Wang Premium
%%Principle \citep{wang2000class}.
%This premium principle is based on quantile regression when several risk factors are considered
%as it can allow a direct estimate of an individual policy's VaR, in which the risk loading is expressed as the difference between the $\tau$-th quantile and $50\%$-th quantile.
%This model explains the needs of risk loading quite well, as it estimates the maximum possible losses that an individual policy may incur with a given probability $1-\tau$ during the forecasting period.
%During the past decades,
%Thus, quantile regression has become an increasingly popular methodology in non-life actuarial science, especially in ratemaking and outstanding claims reserving
%\citep{ kudryavtsev2009using, dong2015risk}.
Compared with GLMs used in the EVPP and SDPP, quantile regression used in VaR Premium Principle shows excellent statistical properties and can better explain the tail characteristics of a response variable's conditional distribution given the values of one or more covariates \citep{koenker1978regression}.
The parameter estimation is less affected by outliers, and the estimation result is more robust \citep{gilchrist2000statistical}.
More recently,
%Following the line of literature,
\citet{heras2018application} propose a so-called Quantile Premium Principle (QPP) based on quantile regression, in which the risk loading is adjusted by a risk loading parameter compared with the VaR Premium Principle.
\cite{baione2019} propose an alternative quantile regression by using a Two-Part Quantile Premium Principle (TSQPP) \footnote{
	In the following, we refer to the model in \cite{heras2018application} as QR and the model in \cite{baione2019} as QRII respectively.
}.
However,
%regardless of the use of quantile regression in various premium principles,
the drawback is that the traditional quantile regression may suffer from the problem of quantile crossing, which may particularly occur at extreme quantiles when the data are rare \citep{dong2015risk}.
To obviate this problem,
\citet{baione2020application} suggest applying a more parsimonious approach to calculate the risk premium based on the parametric quantile regression (PQR) model introduced by  \citet{frumento2016parametric, frumento2017estimating}.
PQR models the regression coefficients as parametric functions of the quantile level, which expands the potential of quantile modelling and avoids the problem of over-parameterization and time-consumption.
%\red{However, regardless of the use of quantile regression and parametric quantile regression, the drawbacks of
%\cite{heras2018application} and \cite{baione2020application}
%is that the QPP used in their studies can not ensure
%the risk premium for individual policies with small risk exposure to be calculated correctly,
%when the actual risk exposure and tariff factor are taken into account (see Section \ref{subsec: Comparison} for more details).
%}
Table \ref{tab: Literature} gives an overview of model specifications and various premium principles discussed in existing classification ratemaking research.
%For comparison, Table \ref{tab: Literature} gives an overview of model specifications and various premium principle discussed in current classification ratemaking research for more details.

%Despite the advantages,
Quantile regression is still quite limiting in the ratemaking process, since the use of VaR as a risk measure is not without criticism \citep{xie2014varying}.
First, it lacks subadditivity, which contradicts the conventional rule of risk diversification.
%For instance,  \citet{kudryavtsev2009using} proposes that the 95\% quantile of an individual policy's aggregate claim amount be used as the risk premium. While it can guarantee that the aggregate claim amount of each policy exceeds its risk premium by no more than 5\%, the probability that whole portfolio's aggregate claim amount exceeds its total risk premium may be less more than 5\%,
%%due to the violation of
%due to risk diversification violation.
%%The total risk of the whole portfolio may be over-estimated.
%\red{In other words, if insurers charge the VaR at 95\% quantile level as the risk premium for individual polices, the total risk premium of the whole portfolio cannot cover its future loss at the same quantile level. }
Second, VaR is insensitive to the magnitude of the loss as it depends only on the probabilities of extreme events but not on their values. This suggests that VaR with a given tail probability may not always be an appropriate risk measure when insurers and regulators are concerned with extreme claims. Third, VaR conveys only a small slice of the loss distribution information as it focuses exclusively on the upper tail of the distribution.
To avoid the aforementioned problems with VaR,
\citet{kuan2009assessing} propose an expectile-based VaR
as an alternative to the VaR as a downside risk measure in the stock market.
The $\tau$-th expectile is defined as the solution to the minimization of asymmetrically weighted mean squared errors, with the weights $\tau$ and $1-\tau$ assigned to
positive and negative deviations, respectively.
%\textcolor[rgb]{1.00,0.00,0.00}{\sout{Therefore, expectile is more sensitive to extreme values of the loss distribution than quantile.}
{Comparing to the $L_1$ norm used in quantile, expectile which utilizes the $L_2$ norm
%is more robust
%to outliers
is more sensitive to extreme values of the
loss distribution.
{Meanwhile, the expectile
%as the loss function
provides
a more smooth surface in the direction of $\tau$, which leads to expectile estimates behaving more stable even for very small or very large values of $\tau$.}
The estimator in expectile is also asymptotically unbiased and normal while the properties of asymptotic unbiasedness and normality are not guaranteed in quantile \citep{waltrup2015expectile}.
Moreover, expectile is determined by tail expectations rather than tail probabilities, thus providing more information about the tails.
Considering expectile as an example of shortfall risk measures is also known as zero-utility premia in the actuarial literature \citep{2002Convex},
expectile can be regarded as a perfectly reasonable alternative to quantile as it depends on both the tail realizations and their probabilities and defines a coherent risk measure for $\tau \ge \frac{1}{2}$ \citep{bellini2014generalized}.
In addition, when dealing with heterogeneous data in regression,
different types of covariates can be introduced in expectile regression models via parametric, semiparametric and nonparametric method which are fully discussed in statistical and financial literature (\cite{kuan2009assessing, yao1996asymmetric, sobotka2013confidence, xie2014varying}}, among others)
{Thus, this regression setting can provide a very nice statistical modelling framework for classification ratemaking in non-life actuarial science.
}

Combining insights from both expectile regression and the actuarial science literature, our main contribution is the design of an efficient premium principle called
{Expectile Premium Principle (EPP)} based on expectile regression, suitable for determining appropriate tariff class risk premium rates \footnote{
We refer to the proposed ratemaking method as ER model.
}.
The EPP is similar to the QPP discussed in \citet{heras2018application} and \citet{baione2020application}, but can satisfy all four properties of a coherent risk measure (i.e., translation invariance, monotonicity, positive homogeneity and subadditivity).
Besides the GLMs and several quantile regressions used as the benchmarks, including QR \citep{heras2018application}, QRII \citep{baione2019}
and PQR models \citep{baione2020application},
we design a simulation study to evaluate the model performance of all competing models.
Based on {a real world} automobile insurance data set, our empirical result shows the proposed method can overcome quantile regression drawbacks and prevent efficiency loss in the estimation of risk premiums.
It also has the advantage of better differentiating the heterogeneity among tariff classes and also has a greater ability to distinguish high risks from low risks.
%Based on a The empirical study also shows that
%we show that
%the proposed method can overcome quantile regression drawbacks and prevent efficiency loss in the estimation of risk premiums.
%in both the simulation and empirical studies.
%Although expectile regression has been found to be applied in various fileds,
To the best of our knowledge, this is the first time that the expectile regression has been implemented into the process of classification ratemaking.

The remainder of the article is structured as follows.
Section \ref{sec: existing-ratemaking} provides a brief description of the standard GLMs and several quantile regressions for classification ratemaking.
Section \ref{sec: ER} describes the proposed Expectile Premium Principle based on expectile regression.
A simulation study is conducted in Section \ref{sec: simulation}.
A numerical application and comparative study are discussed in  Section \ref{sec: results}.
Section \ref{sec: conclusion} concludes this study with suggestions for further research.
The data and R code can be found at \url{https://github.com/lizhengxiao/expectile-regression-risk-loading}.

 \begin{sidewaystable}
	\centering
	\resizebox{\textwidth}{!}{
		\begin{threeparttable}[b]
			\caption{Literature review about risk premium prediction based on various Premium Principles.}
			\renewcommand\arraystretch{4}
			\begin{tabular}{lllllc}
				\toprule
%				\textbf{Classification techniques}		
				 \textbf{Ratemaking models}	
				&	\textbf{Premium Principle}  & \textbf{Risk premium} & \textbf{References} & \textbf{Risk loading} & \textbf{Risk loading parameter}\\
				\hline
				GLMs&{Expected value Premium Principle (EVPP)}  & $\label{EVP} H\left( {{Y}_{i}} \right)=\mathbb{E}\left( {{Y}_{i}} \right)+\varphi \mathbb{E}\left( {{Y}_{i}} \right)$
				&  \citet{heras2018application} & $\varphi \mathbb{E}\left( {{Y}_{i}} \right)$ & $\varphi$\\
				GLMs&{Standard deviation Premium Principle} (SDPP)  & $H\left( {{Y}_{i}} \right)=\mathbb{E}\left( {{Y}_{i}} \right)+\varphi \sqrt{\text{Var}\left( {{Y}_{i}} \right)}$ &\citet{heras2018application}  & $\varphi \sqrt{\text{Var}\left( {{Y}_{i}} \right)}$& $\varphi$\\
				QR*&{VaR Premium Principle} & $H\left( {{Y}_{i}} \right)={{Q}_{{{Y}_{i}}}}\left( \tau  \right)=\inf \left\{ u\in \mathbb{R}:\ {{F}_{{{Y}_{i}}}}\left( u \right)\ge \tau  \right\}$ & \citet{kudryavtsev2009using} & ${{Q}_{{{Y}_{i}}}}\left( \tau  \right) - \mathbb{E}\left( {{Y}_{i}} \right)$& $\tau$\\
				QR&{Quantile Premium Principle} (QPP) & $H\left( {{Y}_{i}} \right)=\mathbb{E}\left( {{Y}_{i}} \right)+\varphi \left[ {{Q}_{{{Y}_{i}}}}\left( \tau  \right)-\mathbb{E}\left( {{Y}_{i}} \right) \right]$ & \citet{heras2018application} & $\varphi \left[ {{Q}_{{{Y}_{i}}}}\left( \tau  \right)-\mathbb{E}\left( {{Y}_{i}} \right) \right]$ & $\varphi (\tau = 95\%)$\\
			QRII&{Two-part quantile Premium Principle} (TSQPP) & $H\left( {{Y}_{i}} \right)=\left( 1-{{p}_{i}} \right){{Q}_{Y_{i}^*}}\left( \tau  \right)$&
				\citet{baione2019}& $\left( 1-{{p}_{i}} \right){{Q}_{Y_{i}^*}}\left( \tau  \right) - \mathbb{E}\left( {{Y}_{i}} \right) $& $\tau$\\
				PQR&{Quantile Premium Principle} (QPP)& $H\left( {{Y}_{i}} \right)=\mathbb{E}\left( {{Y}_{i}} \right)+\varphi \left[ {{Q}_{{{Y}_{i}}}}\left( \tau  \right)-\mathbb{E}\left( {{Y}_{i}} \right) \right]$ & \citet{baione2020application} &
				$\varphi \left[ {{Q}_{{{Y}_{i}}}}\left( \tau  \right)-\mathbb{E}\left( {{Y}_{i}} \right) \right]$
				& $\varphi (\tau = 95\%)$\\
				\bottomrule
			\end{tabular}
			\begin{tablenotes}
				\footnotesize
				\item Notes: Assuming that random variable ${{Y}_{i}}$ denotes the aggregate claim amount for individual policy {$i, i=1,\dots, N$}, the risk premium of policy $i$ can be expressed as a distortion function $H\left( {{Y}_{i}} \right)$ of the random variable ${{Y}_{i}}$.
				\item 			In the EVPP, the risk premium equals the pure premium plus a percentage of the pure premium. $\varphi >0$ denotes the risk loading parameter and $\varphi \mathbb{E}\left( {{Y}_{i}} \right)$ denotes the corresponding risk loading.
				\item In the SDPP, the risk premium equals the pure premium plus a percentage of the standard deviation, $\varphi >0$ denotes the risk loading parameter and $\varphi \sqrt{\text{Var}\left( {{Y}_{i}} \right)}$ denotes the corresponding risk loading.
				\item In the VaR Premium Principle, ${{Q}_{{{Y}_{i}}}}\left( \tau  \right)$ denotes the $\tau$-th quantile of the aggregate claim amount, thus the risk premium is calculated as a quantile of the aggregate claim amount of an individual policy. $\tau$ denotes the risk loading parameter.
				\item In the QPP,
				$\varphi$ is the risk loading parameter, and $\varphi \left[ {{Q}_{{{Y}_{i}}}}\left( \tau  \right)-\mathbb{E}\left( {{Y}_{i}} \right) \right]$ represents the risk loading, which is the difference between the $\tau$-th quantile of the aggregate claim amount and the pure premium. The main difference between the VaR Premium Principle and the QPP is that the risk loading in the QPP is adjusted by risk loading parameter $\varphi$.
				In the QPP, the authors use the pre-determined $\tau=95\%$
				and remain $\varphi$ to be estimated.
				\item In the TSQPP, $1-{{p}_{i}}$ denotes the probability of incurring at least one claim.
				The estimation of risk premium is decomposed between a claim probability that indicates whether the policy has claimed, and a quantile of the claim that has incurred at least one claim $Q_{Y_{i}^*(\tau)}$, where $Y_{i}^*=Y_{i}|Y_{i}>0$.		
				\item In the PQR, parametric quantile regression is used instead of quantile regression model.
			\end{tablenotes}
			\label{tab: Literature}
		\end{threeparttable}
	}
\end{sidewaystable}

\section{The existing ratemaking approaches}\label{sec: existing-ratemaking}
This section gives an overview of current classification ratemaking literature, including GLMs and three quantile regression models.

Suppose an insurance portfolio contains $N$ policies, ${{R}_{i}}$ indicates whether or not policy $i$ has a claim submitted, ${{Y}_{i}}$ represents its aggregate claim amount, ${{w}_{i}}$ denotes its exposure, and ${{\bm{x}}_{i}}$ stands for a vector of covariates $(i=1,\cdots ,N)$.

\subsection{Risk premium estimation via GLMs}
To set the premium for individual policyholders, it is a common practice to separate claim probability and non-zero aggregate claim amount,
in which the claim probability component models the probability that a given policy incurs claims, and the non-zero component models the aggregate claim amount given that at least one claim has been incurred \citep{frees2009regression,frees2013actuarial}.
Such a decomposition method is commonly called the two-part GLMs framework.
In two-part GLMs, for claim probability we assume that $R_i$ follows the binomial distribution with parameter $1-p_i$, and consider the conventional logistic regression model:
\begin{eqnarray}\label{logit}
\text{logit}\left[ \frac{1-{{p}_{i}}}{{{w}_{i}}} \right]=\bm{x}'_{i}\bm{\alpha },
\end{eqnarray}
where
$R_i=0$ denotes the policy having no claims with probability $p_i$,
$\text{logit}(t)=\log\left( {t}/{(1-t)}\right)$ is the logit function, $\bm{x}_{i}=(1,x_{i1},\cdots ,x_{ik})$ represents the $(k+1)$-dimensional vector of covariates, and $\bm{\alpha }=({{\alpha }_{0}},{{\alpha }_{1}},\cdots ,{{\alpha }_{k}}{)}$ denotes the corresponding regression coefficients to be estimated.
The left-hand side of (\ref{logit}) is the log odds ratio per exposure, and is corrected for risk exposure ${{w}_{i}}$ \citep{de2008generalized}. Correspondingly, the probability of at least one claim occurring can be obtained by $1-{{p}_{i}}={{w}_{i}}\frac{\exp (\bm{x}'_{i}\bm{\alpha })}{1+\exp (\bm{x}'_{i}\bm{\alpha })}$.
%One of the main advantages of this predictive technique is its interpretability, by capturing the non-linear relationship between claim probability and the risk factors.
For the non-zero aggregate claim amount, we employ Gamma (GA) regression to model the skewness and heavy tail.
The log link function is considered
\citep{kudryavtsev2009using},
%Considering the log link function is quite popular in ratemaking process, since it is well connected with the multiplicative framework \citet{mack1997schadenversicherungsmathematik,kudryavtsev2009using},
and the expected claim amount can be obtained by
\begin{equation}\label{log-link}
\log ({{\mu }_{i}})=\bm{x}'_{i}\bm{\beta },
\end{equation}
where $\bm{\beta }=({{\beta }_{0}},{{\beta }_{1}},\cdots ,{{\beta }_{k}}{)}$ denotes the corresponding regression coefficients to be estimated in the non-zero claim amount component.

The pure premium of policy $i$ in the two-part GLMs is expressed as the expected value of the aggregate claim amount, which is given by
\begin{equation}
\mathbb{E}\left[ {{Y}_{i}} |\bm{x}_i \right]=\left( 1-{{p}_{i}} \right){{\mu }_{i}}
\end{equation}
thus resulting in the risk premium by applying the EVPP and
SDPP given respectively by:
\begin{align}
H\left( {Y}_{i} |\bm{x}_i \right)&=\mathbb{E}\left( {{Y}_{i}} |\bm{x}_i \right)+\varphi \mathbb{E}\left( {{Y}_{i}}  |\bm{x}_i\right),\\
H\left( {{Y}_{i}}|\bm{x}_i  \right)&=\mathbb{E}\left( {{Y}_{i}}  |\bm{x}_i\right)+\varphi \sqrt{\text{Var}\left( {{Y}_{i}}  |\bm{x}_i\right)},
\end{align}
where $\varphi$ is the risk loading parameter.

\subsection{Risk premium estimation via QR, QRII and PQR}
\label{subsec: models}
Similar to the two-part GLMs, an alternative method for modelling the non-zero aggregate claim amount is to apply quantile regression. As aforementioned,
three models (i.e., QR, QRII and PQR) are discussed in existing literature, where two different premium principles (i.e., QPP and TSQPP) are employed, see Table \ref{tab: Literature} for more details.
%two quantile regressions (i.e., QR and PQR) and two premium principles (i.e., QPP and TSQPP) are employed.

For QR model, we follow the general guidelines for classification analysis in the risk premium prediction context as proposed in \citet{heras2018application} based on QPP,
 thus resulting in the risk premium of policy $i$ simply given by
\begin{equation}\label{twp-part quantile}
	H\left( {{Y}_{i}} |\bm{x}_i\right)=\mathbb{E}(Y_i|\bm{x}_i)+\varphi \left[ {{Q}_{{{Y}_{i}}}}\left( \tau  |\bm{x}_i\right)-\mathbb{E}\left( {{Y}_{i}} |\bm{x}_i\right) \right],
\end{equation}
where ${{\bm{x}}_{i}}$ stands for a vector of covariates,
$\mathbb{E}(Y_i|\bm{x}_i)$ is the pure premium obtained by GLMs,
and
${{Q}_{{{Y}_{i}}}}\left( \tau  |\bm{x}_i\right)$ represents the VaR with $\tau$-level.
\eqref{twp-part quantile} refers to the Quantile Premium Principle (QPP), in which the individual risk premium is a weighted average of pure premium and the $\tau$-level quantile of $Y_i$.
In \citet{heras2018application}, $\tau$ is set to be $95\%$ and $\varphi$ is denoted by the risk loading parameter.
In order to estimate the $\tau$-quantile of $Y_i$, we can estimate the $\tau_i^*$-th quantile of $Y_{i}^{*}$,
where  $Y_{i}^{*}={{Y}_{i}}|{{Y}_{i}}>0$ represents the non-zero aggregate claim amount given that policy $i$ incurs at least one claim,
with the following relationship: $\tau _{i}^{*}=({\tau -{{p}_{i}}})/({1-{{p}_{i}}})$
\footnote{
It can be easily proved that $Q_{Y_i}(\tau|\bm{x}_{i})=Q_{Y_{i}^{*}}(\tau _{i}^{*}|\bm{x}_{i})$
}
.
This ratemaking framework refers to a two-part quantile regression because the probability of having no claim   is estimated using the logistic regression model; the non-zero claim amount is modeled by quantile regression.
Considering the log link function is quite popular as it is well connected with the multiplicative framework,
the covariates can be introduced into log-transformed quantile as follows:
\begin{equation}
\log{{Q}_{Y_{i}^{*}}}\left( {{\tau_{i}^{*}}}\left| \bm{x}_{i}\right. \right)=
%\exp\left(
\bm{x}'_{i}{{\bm{\gamma }}^{\tau_{i}^{*}}}.
%\right).
\end{equation}
Here we refer ${{Q}_{Y_{i}^{*}}}\left( {{\tau_{i}^{*}}}\left| \bm{x}_{i}\right. \right)$ to the VaR at quantile level ${\tau_{i}^{*}}$.
In their studies, the response variable is the log-transformed non-zero aggregate claim amounts of individual policies that submit at least one claim.
The vectors of regression coefficients $\bm{\gamma }^{\tau_{i}^{*}}=(\gamma _{0}^{\tau_{i}^{*}},\gamma _{1}^{\tau_{i}^{*}},\cdots ,\gamma _{k}^{\tau_{i}^{*}})$ are not the same for tariff classes because of their different quantile levels,
which indicates its application requires the calibration of a number of quantile regression models on different quantile levels equal to the number of risk classes.
%used for the frequency model which,
%in large portfolios, could be equal to thousands or even millions.
Regression coefficients estimation can be derived by solving the following minimization problem with R package \texttt{quantreg}: Quantile Regression; see \citet{koenker2001quantile} for more details.
%However, this approach is questionable from a
%practical point view as it requires the calibration of a number of quantile regression models on different quantile levels, which may cause the problems of over-parameterization and heavy computation.

PQR as the second model considered in our analysis expresses the regression coefficients as some parametric functions of the quantile level \citep{baione2020application}.
This model has several advantages, including parsimony, efficiency, and simple interpretation. PQR results are associated with the choice of the quantile level function. In practice, the choice of the function must ensure that the quantile is monotonically increasing.
{For instance}, polynomials, splines, trigonometric functions, and quantile functions of standard normal distributions could be used. Here we choose a polynomial function to capture the relationship between quantile levels and the coefficients of the quantile regression model. The covariates can be introduced into a log-transformation quantile given by
\begin{align}
\log{{Q}_{Y_{i}^{*}}}\left( {\tau_{i}^{*}}|\bm{x}_{i} \right) &=\bm{x}_{i}'\bm{\gamma }\left( {\tau_{i}^{*}}\left| \bm{\theta } \right. \right),\nonumber\\
{{\gamma }_{j}}( {\tau_{i}^{*}}|\bm{\theta})&={{\theta }_{0j}}+{{\theta }_{1j}} {\tau_{i}^{*}}+{{\theta }_{2j}}{\tau_{i}^{*}}^{2},\quad j=0,1,\cdots ,k,
\label{eq: PQR}
\end{align}
where $\bm{\gamma }\left( {\tau_{i}^{*}}\left| \bm{\theta } \right. \right)={{\left( {{\gamma }_{0}}\left( {\tau_{i}^{*}}\left| \bm{\theta } \right. \right),\cdots ,{{\gamma }_{k}}\left( {\tau_{i}^{*}}\left| \bm{\theta } \right. \right) \right)}}$ denotes the corresponding vector as a function of quantile level $ {\tau_{i}^{*}}$ conditioning on finite-dimensional parameters $\bm{\theta }$,
and ${{\gamma }_{j}}\left( {\tau_{i}^{*}} \right)$ is a polynomial function.
We also refer ${{Q}_{Y_{i}^{*}}}\left( {\tau_{i}^{*}}|\bm{x}_{i} \right)$ to the VaR at ${\tau_{i}^{*}}$-level.
The estimation procedure can be implemented with R package \texttt{qrcm}: quantile regression coefficients modelling; see \citet{gilchrist2000statistical} and \citet{frumento2016parametric, frumento2017estimating} for more details.
The advantage of this model is that it enables the estimation of the conditional quantile of the response variable, given a set of covariates, for each quantile level in a single run.
%This model is also analyzed in \citet{baione2020application}.

The third model is an alternative quantile regression model  (QRII) based on TSQPP proposed by \citet{baione2019}, resulting in the risk premium of policy $i$
simply given by
\begin{equation}\label{twp-part quantile2}
H\left( {{Y}_{i}|\bm{x}_i}\right)=\left( 1-{{p}_{i}} \right){{Q}_{Y_{i}^{*}}}\left( \tau \left| {{\bm{x}}_{i}}\right. \right),
\end{equation}
where the quantile level $\tau$ denotes the risk loading parameter and ${{Q}_{Y_{i}^{*}}}\left( \tau \left| {{\bm{x}}_{i}} \right. \right)$
denotes the VaR of non-zero aggregate claim amount $Y_{i}^{*}$ at $\tau$-level.
The risk premium calibration is assessed by means of a conventional quantile regression model using a single quantile level for each risk class. However, the drawback of this approach is that it is not possible to obtain a unique quantile level of the aggregate claim amount for each tariff class by fixing a single quantile level $\tau$.
%Note that the risk loading parameter $\tau$ in the two-part quantile premium principle is fixed for all individual policies.

\section{The proposed ratemaking approach}\label{sec: ER}
%In this study, we consider the expectile quantile (ER) regressions as an alternative to the quantile regressions,
%to model the aggregate claim amount $Y_i$ for $i=1,...,N$.
%The expectile regression model is first proposed in
\subsection{General model for expectile regression (ER)}
To overcome the drawbacks of quantile regression, it is possible to adopt expectile regression to model the aggregate claim amount $Y_i$ for $i=1,...,N$ based on a set of covariates (or risk factors) $\bm{x}_i$.
%The expectile
%The term "expectile" has probably been suggested as a combination
%of expectation and quantile"
%Similar to the definition of VaR, the $\tau$-level expectile is understood as the maximal possible loss within a given holding period under the
%expectile level $1 - \tau$. Since expectile is also a VaR, it is also the maximal possible loss within a given holding period under the quantile
%$1-\alpha^*$, where $1-\alpha^*$ is the ex post tail probability associated with the expectile.
%
%Based on a large enough number $N$ of observations $(y_i, \bm{x}_i), i=1,...,N$,
%expectile regression for the $\tau$-expectile with $\tau\in (0,1)$ relies on the regression specification
%\begin{equation}
%y_{i} = \eta_{i\tau} + \epsilon_{i\tau}, \quad i = 1,...,N,
%\end{equation}
%where $\eta_{i\tau}$ is a (expectile-specific) predictor and $ \epsilon_{i\tau}$ are independent error terms. Instead of imposing the usual mean regression model assumption that
%$\mathbb{E}( \epsilon_{i\tau})=0$, expectile regression relies on the assumption that for the expectile function
%$v$
%holds that $v_{\epsilon_{i\tau}}(\tau)=0$.
%This implies that the conditional expectile of the response $y_i$ is given by the predictor $\eta_{i\tau}$.
%They are implicitly defined by $v_{\epsilon_{i\tau}}(\tau)=\arg \min_{m}\mathbb{E}\left[\omega_{i\tau}(m, \epsilon_{i\tau})(\epsilon_{i\tau}-m)^2\right]$
%
We let the $\tau$-th conditional expectile ${{v}_{Y_{i}}}\left( {\tau}\left| \bm{x}_{i}\right. \right)$ of $Y_{i}$ be modeled by the linear specification $\bm{x}_{i}'{{\bm{\gamma }}^\tau}$
\begin{equation}
	\label{eq: er}
{{v}_{Y_{i}}}\left( {\tau}\left| \bm{x}_{i}\right. \right)=\bm{x}_{i}'{{\bm{\gamma }}^\tau},
\end{equation}
where $\bm{\gamma }^\tau=(\gamma _{0}^\tau,\gamma _{1}^\tau,\cdots ,\gamma _{k}^\tau)$ denotes the corresponding regression coefficients to be estimated at $\tau$-level \footnote{
Note that (positive) aggregate claim amount $Y_i^*$ is modelled as a response via log link function in quantile regression model,
%is modelled as a response variable in quantile regression model,
while the aggregate claim amount $Y_i$ is modelled as a response via identical link function (linear form) in expectile regression model}.
%It is well know that the relation $\tau=\Pr\left[Y_{i}\le{{v}_{Y_{i}}}\left( {\tau}\left| \bm{x}_{i}\right. \right)|\bm{x}_{i}\right]$ holds.

In statistical practice, the distribution-free approach is often used for estimation \citep{newey1987asymmetric}.
%It is based on a large enough number $N$ of observations $(y_i, \bm{x}_i), i=1,...,N$.
%In this framework,
For a given value of $\tau$,
the estimates $\hat{\bm{\gamma}}^{\tau}$ from \eqref{eq: er}
can be obtained by minimizing the asymmetric least squares function
\begin{equation}
	\label{eq: weighted-loss-function}
	\min_{\bm{\gamma}^{\tau}}\sum_{i=1}^{N}
	\left[
	\abs{\tau - I_{\left\{y_{i}-\bm{x}_i'\bm{\gamma}^{\tau}\le 0\right\}}}\cdot \abs{y_i-\bm{x}_i'\bm{\gamma}^{\tau}}^2
	\right],
%	\left[
%	\omega_{i\tau}
%	\cdot \abs{y_i-\bm{x}_i\bm{\gamma}^{\tau}}^2
%	\right],
\end{equation}
%with respect to $\bm{\gamma}^{\tau}$,
where
$\tau\in\left[0,1\right]$ determines the degree of asymmetry of the
loss function.
If $\tau=0.5$, \eqref{eq: weighted-loss-function} reduces to the standard least-squares objective function, and $v_{Y_i}(0.5|\bm{x}_i)$ is just the conditional mean of $Y_i|\bm{x}_i$.
It is known as the mean regression.
%This model is more popular in practice than the more general approach of expectile regression. Nevertheless, the more general approach is much more important for actuarial practice.

According to \citet{newey1987asymmetric} and \citet{sobotka2013confidence} among others, the estimates can be derived fairly easily based on iteratively
reweighted least squares (IRLS), that is, as the solution to the equation
\[
{\bm{\gamma}}^{\tau}=\left[\sum_{i=1}^N	\abs{\tau - I_{\left\{y_{i}-\bm{x}_i'{\bm{\gamma}}^{\tau}\le 0\right\}}} \bm{x}_i \bm{x}'_i\right]^{-1}\sum_{i=1}^N	\abs{\tau - I_{\left\{y_{i}-\bm{x}_i'\bm{\gamma}^{\tau}\le 0\right\}}}\bm{x}_i {y}_i.
\]
%Another attractive feature of expectile regression is that the expectile regression estimator depends on the entire distribution's shape, while quantile regression estimator only relies on the percentiles of the estimated tail distribution. Hence, the expectile regression estimator contains additional information about tail distribution magnitude and reflects the real value more accurately, especially for heavy-tailed distributions used in non-life actuarial science.
%Thus, quantile regrssion is more resisitant to outliers than expectile regression as quantile regression utilizes the $L_1$ norm.

%The least asymmetrically weighted squares estimate is asymptotically normal, i.e.

The asymptotic distribution of $\hat{\bm{\gamma}}^{\tau}$ is given by
\[
\sqrt{N}\left(\hat{\bm{\gamma}}^{\tau} - {\bm{\gamma}}^{\tau}\right) \stackrel{D}{\longrightarrow}\mathbf{\text{N}}\left(\bm{0},
\bm{W}^{-1}\bm{V}\bm{W}^{-1}\right)
\]
where
\begin{align*}
&\bm{W}=\mathbb{E}\left[w_i(\tau) \bm{x}_i\bm{x}'_i\right], \\
&\bm{V}=\mathbb{E}\left[w_i^2(\tau) u_i^2(\tau)\bm{x}_i\bm{x}'_i\right],
\end{align*}
with the residuals $u_i(\tau)=y_i-\bm{x}'_i\bm{\gamma}^\tau$, and the weights $w_i(\tau)=\left|\tau-I_{\left\{u_i(\tau)<0\right\}}\right|$. $\bm{V}$ is partitioned conformably with $\bm{\gamma}^{\tau}$.
%In order to construct large sample confidence intervals, a consistent estimator of the asymptotic must be constructed.
The asymptotic covariance matrix of the vector $\bm{\gamma}^{\tau}$ can be constructed using the residuals $\hat{u}_i(\tau)$ and the estimated weights $\hat{w}_i(\tau)=\left|\tau-I_{\left\{\hat{u}_i(\tau)<0\right\}}\right|$, which is given by
\[
\widehat{\text{Var}}\left(\bm{\gamma}^{\tau}\right) = \hat{\bm{W}}^{-1}\hat{\bm{V}}\hat{\bm{W}}^{-1}.
\]
where
\begin{align*}
\hat{\bm{W}}&=\sum_{i=1}^N\hat{w}_i(\tau)\bm{x}_i\bm{x}'_i\big/N,\\
\hat{\bm{V}}&=\sum_{i=1}^N\hat{w}_i^2(\tau)\hat{u}_i^2(\tau)\bm{x}_i\bm{x}'_i\big/N.
\end{align*}
The square root of the diagonal elements of $\bm{D}\left(\bm{\hat{\gamma}}^\tau\right)=\hat{\bm{W}}^{-1}\hat{\bm{V}}\hat{\bm{W}}^{-1}$ yield the standard
errors of the estimates of the parameters, leading to asymptotic $100(1- p)\%$
confidence intervals for $\gamma_{j}^\tau$ for $j = 0,1,...,k$ given by:
\[
\left(\gamma_{j}^\tau - z_{q/2}\sqrt{D_{jj}(\bm{\hat{\gamma}}^\tau)}, \gamma_{j}^\tau + z_{q/2}\sqrt{D_{jj}(\bm{\hat{\gamma}}^\tau)}\right),
\]
where $z_{q/2}$ denotes the upper $q/2$ quantile of the standard normal distribution.
The parameter estimation procedure of expectile regression model is implemented with R package \texttt{expectreg}: expectile regression.

\subsection{The Expectile Premium Principle}

As a reasonable alternative risk measure, we propose using the distance between some pre-established $\tau$-th expectile of the aggregate claim amount, $v_{Y_i}(\tau|\bm{x}_i)$
and its expected value, $\mathbb{E}(Y_i|\bm{x}_i)$, to calculate the risk loading.
The resulting risk premium is calculated as a convex combination of $v_{Y_i}(\tau|\bm{x}_i)$  and $\mathbb{E}(Y_i|\bm{x}_i)$:
\begin{equation}\label{eq: EPP}
H\left( {{Y}_{i}} |\bm{x}_i\right)=\mathbb{E}(Y_i|\bm{x}_i)+\varphi \left[ {{v}_{{{Y}_{i}}}}\left( \tau  |\bm{x}_i\right)-\mathbb{E}\left( {{Y}_{i}} |\bm{x}_i\right) \right],
\end{equation}
where $\varphi$ is risk loading parameter and ${{v}_{{{Y}_{i}}}}\left( \tau  |\bm{x}_i\right)$ can be estimated by applying expectile regression.
The proposed premium principle has some advantages over the QPP used in
\eqref{twp-part quantile},
as for $\tau\ge\frac{1}{2}$ , ${{v}_{{{Y}_{i}}}}\left( \tau  |\bm{x}_i\right)$ is a coherent risk measure because it satisfies the well-known axioms introduced by \citet{artzner1999coherent}. Indeed, it is easy to see that
%it can be easily shown that this premium principle satisfies all the properties of coherent risk measure:
\begin{itemize}
	\item $H\left( {{Y}_{i}}+h |\bm{x}_i\right)=H\left( {{Y}_{i}} |\bm{x}_i\right)+h$, for $h\in \mathbb{R}$ (translation invariance)
	\item $U_{i} \le Y_{i}$ a.s. $\Rightarrow H\left( {{U}_{i}} |\bm{x}_i\right)\le H\left( {{Y}_{i}} |\bm{x}_i\right)$  (monotonicity)
	\item $H\left( \lambda {{Y}_{i}} |\bm{x}_i\right)=\lambda H\left( {{Y}_{i}} |\bm{x}_i\right)$, for $\lambda \ge 0$ (positive homogeneity)
	\item $H\left(U_i + {{Y}_{i}} |\bm{x}_i\right) \le H\left({{U}_{i}} |\bm{x}_i\right) + H\left({{Y}_{i}} |\bm{x}_i\right)$ (subadditivity).
\end{itemize}
Thus, we will call it the Expectile Premium Principle (EPP)\footnote{In the following simulation and empirical study, we use the pre-determined expectile level $\tau=95\%$
	and the risk loading parameter $\varphi$ remains to be estimated.}.

The expectile in \eqref{eq: EPP}  can also be interpreted an average that balances between
a conditional downside mean and conditional upside mean \citep{kuan2009assessing}
that is,
\[
v_{Y_i}(\tau|\bm{x}_i) = \eta \mathbb{E}\left[{Y_i|\bm{x}_i,Y_i>v_{Y_i}(\tau)}\right] + (1-\eta)\mathbb{E}\left[{Y_i|\bm{x}_i,Y_i\le v_{Y_i}(\tau)}\right],
\]
where $\eta= \tau \left[1-F_{Y_i}(v_{Y_i}(\tau))
\right]/\left\{
\tau \left[1-F_{Y_i}(v_{Y_i}(\tau))
\right] +  (1-\tau) F_{Y_i}(v_{Y_i}(\tau))\right\}
$.
This property distinguishes the expectile
from the expected shortfall because the former is determined both by the upper and lower tails of distribution.
The level $\tau$ can be understood as the relative cost of the expected margin shortfall.
Moreover,
the proposed QPP is a more conversational risk measure for the insurers and regulators, because for large $\tau$,
expectiles are more conservative than the usual quantiles \citep{bellini2014generalized}.

\section{Simulation study}\label{sec: simulation}
To investigate what the optimal ratemaking methods for
predicting the risk premium for individual polices in
non-life insurance are, we performed a simulation study where different values of the true risk loading are considered.
%where
%the characteristics for the loss distribution
%\red{the characteristics for the risk loading
%is varied.}
All simulations and calculations are performed with the help of R version 4.0.2.

The simulation study is based on a series of generated  data sets.
Each data set consists of $N=5,000$ observations of aggregate claim amount $Y_i$ and three covariates $\bm{x}_i=(1,x_{i1},x_{i2},x_{i3})$ for $i=1,...,N$.
%\textcolor[rgb]{1.00,0.00,0.00}{\sout{with $k = 3$}}.
%\textcolor[rgb]{1.00,0.00,0.00}{\sout{The aggregate claim amount} $Y_i$}
%and $Y_i$
$Y_i$ is generated from the Tweedie distribution with the mean $\mu_i=\exp(\bm{x}'_i\bm{\beta})$
and variance $\phi\mu_i^p$,
where $\phi$ is the dispersion parameter and $p$ is the power parameter.
We consider three binary variables $x_{i1},x_{i2},x_{i3}$ being generated from the Bernoulli  distribution with probability
$\bm{\theta} = (0.5,0.6,0.8)$,
%$0.5,0.6$ and $0.8$,
and
the regression coefficient $\bm{\beta}=(5, 0.5, 0.5, 0.5)$.
This results in a total number of 8 tariff classes in one simulated data set.
For the sake of brevity, we set the risk exposure $w_i=1$ for all observations.
Thus, the true pure premium of each observation can be denoted by $\mathbb{E}(Y_i|\bm{x}_i)=\mu_i$ for $i=1,...,N$.
Following \cite{heras2018application} approach, we also assume
that the true risk premium of each observation can be expressed as $\mu_i+\varphi \mu_i$, resulting in the true total risk premium of the whole portfolio being
$C=\sum_{i=1}^{N}(\mu_i+\varphi \mu_i$) with
different values of $\varphi$ from 0 to 0.15.
A larger value of $\varphi$ indicates higher risk loading.
Given the fact that the aggregate claim amount data is usually highly right-skewed with a point mass at zero in non-life insurance, the power parameter in Tweedie distribution is assumed to be $p=1.65$ and the dispersion parameter $\phi = 120$.

In order to estimate risk premium rates for 8 tariff classes, the first step is to use two-part GLMs to obtain the estimate of pure premium for individual policies. Then, we apply the ratemaking methods discussed in Section \ref{sec: existing-ratemaking}  and Section \ref{sec: ER}  to obtain
the risk loading prediction in QR, PQR, QRII and ER models.
{Specifically, the risk loading parameter used in various premium principle must be pre-determined
by distributing the total risk premium
$C$ to individual policies by solving the following equation \footnote{
%The total risk premium is computed as the sum of true risk premium for all individual polices, that is, $C=\sum_{i=1}^{N}\left(\mu_i+\varphi \mu_i\right)$.
For model comparison, we control the total risk premium of simulated insurance portfolios.
This method refers to the premium allocation method, which is also discussed in \cite{heras2018application}.
Thus, we can obtain the estimates of risk loading parameters and regression coefficients in QR, PQR, QRII and ER models in every simulation run.
The definition of the risk loading parameter in QR, PQR, QRII and ER models can be found in Table \ref{tab: Literature}.
%the expectile level in ER,
%the quantile level in QRII, the ris
%and risk loading parameters in QR, QRII and PQR by using this premium allocation method in every simulation run.
%	and only allow $C$ to be varying from the values of risk loading parameter $\varphi$.
%A larger value of $\varphi$ indicates higher risk loadings.
}:
}

%used in QR, QRII, PQR and ER, as well as the estimates of regression coefficients can be easily obtained

%by distributing the total risk premium
%$C$ to individual policies by solving the following equation \footnote{
%%The total risk premium is computed as the sum of true risk premium for all individual polices, that is, $C=\sum_{i=1}^{N}\left(\mu_i+\varphi \mu_i\right)$.
%{For model comparison, we control the total risk premium of simulated insurance portfolios.
%%	and only allow $C$ to be varying from the values of risk loading parameter $\varphi$.
%%A larger value of $\varphi$ indicates higher risk loadings.
%Thus, we can obtain the expectile level in ER and risk loading parameters in QR, QRII and PQR by using this premium allocation method in every simulation run.}
%}:
%\citep{heras2018application, baione2020application}:
\begin{equation}
	\sum_{i=1}^{N}{H\left( {{Y}_{i}}|\bm{x}_i,w_i = 1\right)}=C,
	\label{eq: allocation}
\end{equation}
where $H\left( {{Y}_{i}} |\bm{x}_i,w_i = 1\right)$ denotes the risk premium prediction with the risk exposure set to 1 for the premium priciple by using QR, PQR, QRII and ER models.

%For each parameter setting, we sample $N=5,000$ observations from the aforementioned Tweedie distribution, and then consider the following measures for the estimated risk premium.
%For a
%
%the bias and
%the mean square error (MSE) to evaluate the model performance for each tariff class.
%We repeat this procedure $ T=  5,000$ times.

For each parameter setting, we repeat this procedure $ T=  5,000$ times
%generated $T = 5,000$ data sets from the aforementioned Tweedie distribution
and
compute the mean square error (MSE) to evaluate model performance.
Let $t=1,...,T$ be the number of data sets, $j=1,...,J$  be the number of ratemaking models (i.e., QR, PQR , QRII and ER models with $J=4$),
$s=1,...,S$ be the number of tariff classes with $S=8$,
$z_{s}^{(t)}$ and $\hat{z}_{sj}^{(t)}$ be the true risk premium and the predicted risk premium for $s$-th tariff class {in $t$-th simulated data set using $j$-th ratemaking model}.
%$\bar{z}_{sj} = \frac{1}{T}\sum_{t=1}^{T}z_{s}^{(t)}$ be the mean estimated risk premium for for $s$-th tariff class.
%The MSE of $j$-th model is defined as
%\begin{equation}
%\text{MSE}_j := \mathbb{E}\left[\frac{1}{S}\sum_{s=1}^{S}(z_{sj} - \hat{z}_{sj})^2   \right].
%\label{eq: mse}
%\end{equation}
In the $t$-th simulated data set, we obtain an estimate of MSE of the $j$-th model \citep{kramer2013total} via
\begin{equation*}
	\widehat{\text{MSE}}_j^{(t)} := \frac{1}{S}\sum_{s=1}^{S}(z_{s}^{(t)} - \hat{z}_{sj}^{(t)})^2.
\end{equation*}
Then, we compare the mean MSE of the $j$-th model
%The quality of the estimated risk premium is evaluated by
\begin{equation}
\label{eq: barmse}
	\overline{\text{MSE}}_j := \frac{1}{T}\sum_{t=1}^{T}\widehat{\text{MSE}}_j^{(t)}
\end{equation}
computed over all $T$ simulation runs. Note that its variance can be estimated via
\begin{equation*}
S_{\overline{\text{MSE}}_j}^2=\frac{1}{T}\cdot\frac{1}{T-1}\sum_{t=1}^{T}\left(\widehat{\text{MSE}}_j^{(t)} - \overline{\text{MSE}}_j \right)^2.
\end{equation*}
In addition, we are also interested in the bias, sample variance and MSE of the risk premium estimated in each tariff class, which is defined as
\begin{equation}
\label{eq: hatmse}
\widehat{\text{Bias}}_{sj}:= \frac{1}{T}\sum_{t=1}^{T}\left(z_{s}^{(t)} - \hat{z}_{sj}^{(t)}\right),\quad \quad
\widehat{\text{MSE}}_{sj} := \frac{1}{T}\sum_{t=1}^{T}\left(z_{s}^{(t)} - \hat{z}_{sj}^{(t)}\right)^2,
\end{equation}
and the sample variance
\begin{equation}
\label{eq: samplevar}
S_{{sj}}^2=\widehat{\text{MSE}}_{sj} + \widehat{\text{Bias}}_{sj}^2, \quad \text{for} \quad s = 1,...,S.
\end{equation}

	\begin{table}[!h]
	\small
	%\centering
	\renewcommand\arraystretch{1.3} % µ÷ÕûÐÐ¸ß¶È
	\caption{Simulation procedures.}
	\setlength{\tabcolsep}{13mm}%{   % ¼õÉÙ±í¸ñµÄÁÐ¼ä¾àÀë
	\setlength{\parindent}{-2em}     % Ê×ÐÐËõ½ø
	%\hangafter 1
	%\hangindent 2.5em
	% Table generated by Excel2LaTeX from sheet 'mse.all'
	\begin{tabular}{m{16cm}}%{l}
		\toprule
		\setlength{\parindent}{-2em}
		\textbf{Input:} $\bm{\theta}$,\,$\bm{\beta}$,\,$\phi$,\,$p$,\,$\varphi$,\,$N$,\,$T$,\,$J$,\,$S$.\\
		\setlength{\parindent}{-2em}
		\textbf{Output:} $S^2_{sj}$, $\widehat{\text{MSE}}_{sj}$ and $\overline{\text{MSE}}_j$ for $s=1,...,S (S=8)$ and $j=1,...,J (J=4)$.\\
		\setlength{\parindent}{-2em}
		\textbf{For $t=1$ to $T$ do}.\\
		%\begin{tabular}{m{14cm}}	
		\setlength{\parindent}{-2em}
		\textbf{(1)} Generated three covariates $x_{i1}^{(t)},x_{i2}^{(t)},x_{i3}^{(t)}$ from the Bernoulli distribution with probability $\bm{\theta} = (0.5,0.6,0.8)$ for $i=1,...,N$. \\
		%\end{tabular}\\
		%			Generated three covariates from $x_i$: $x_i\sim B(\theta)$, $\theta = c(0.5,0.6,0.8)$. \\
		\setlength{\parindent}{-2em}
		\textbf{(2)} Generated the aggregate claim amount $Y_i^{(t)}$ from the Tweedie distribution $(\mu_i^{(t)}, p, \phi)$ with $\mu_i^{(t)}=\exp(\bm{x}_i^{(t)}{'} \bm{\beta})$, $\phi = 120$ and $p=1.65$, where $\bm{x}_i^{(t)}=(1,x_{i1}^{(t)},x_{i2}^{(t)},x_{i3}^{(t)})$ and $\bm{\beta}=(5,0.5,0.5,0.5)$, for $i=1,...,N$. \\
		\setlength{\parindent}{-2em}
		\textbf{(3)} Calculated the true risk premium of individual policies $z_i^{(t)}$ and of tariff classes $z_s^{(t)}$, where $z_i^{(t)}=\mu_i^{(t)}+\varphi\mu_i^{(t)}$ and $z_s^{(t)}=\mu_s^{(t)}+\varphi\mu_s^{(t)}$
		for $i=1,...,N$ and $s=1,...,S$ with values of  $\varphi\in [0,0.15]$.\\
		%		   \qquad which is defined as $z_i^{(t)}=\mu_i^{(t)}+\varphi\mu_i^{(t)}$,\, $\varphi\in [0,0.15]$. \\
		\setlength{\parindent}{-2em}
		\textbf{(4)} Generated the true total risk premium of the whole portfolio: $C^{(t)}=\sum_{i=1}^{N} z_i^{(t)}$. \\
		\setlength{\parindent}{-2em}
		\textbf{(5)} Estimated the pure premium $\hat{\mu}_i^{(t)}$ and $\hat{\mu}_s^{(t)}$, the probability  $\hat{p}_i$ and $\hat{p}_s$ using two-part GLMs for $i=1,...,N$ and $s=1,...,S$. \\
		\setlength{\parindent}{-2em}
		\textbf{(6)} Estimated the risk loading parameters and the regression coefficients in QR, PQR, QRII and ER models based on premium allocation method, thus resulting in the risk premium prediction
		$\hat{z}_{sj}^{(t)}$ for $s=1,...,S$ and $j=1,...,J$. Specifically,\\
		\setlength{\parindent}{-1.5em}
		\textbf{(6.1)} In QR model,  estimated the risk loading parameter $\hat{\varphi}_{\text{QR}}^{(t)}$ by using \eqref{twp-part quantile}  and \eqref{eq: allocation}, and
		$\hat{z}_{s1}^{(t)}=\hat{\mu}_s^{(t)} + \hat{\varphi}_{\text{QR}}^{(t)}\left[Q_{Y_i^{(t)}}(0.95|\bm{x}_s^{(t)}) - \hat{\mu}_s^{(t)} \right].$ 	\\
		\setlength{\parindent}{-1.5em}
		\textbf{(6.2)} In PQR model,  estimated the risk loading parameter $\hat{\varphi}_{\text{PQR}}^{(t)}$ by using \eqref{eq: PQR}  and \eqref{eq: allocation}, and
		$\hat{z}_{s2}^{(t)}=\hat{\mu}_s^{(t)} + \hat{\varphi}_{\text{PQR}}^{(t)}\left[Q_{Y_i^{(t)}}(0.95|\bm{x}_s^{(t)}) - \hat{\mu}_{s2}^{(t)} \right].$	\\
		\setlength{\parindent}{-1.5em}
		\textbf{(6.3)} In QRII model, estimated the risk loading parameter $\hat{\tau}_{\text{QRII}}^{(t)}$ by using (\ref{twp-part quantile2}) and (\ref{eq: allocation}), and
		$\hat{z}_{s3}^{(t)}=\left( 1-{\hat{p}_{s}^{(t)}} \right){{Q}_{Y_{i}^{*(t)}}}\left( \hat{\tau}^{(t)}_{\text{QRII}} \left| {{\bm{x}}_{i}^{(t)}}\right. \right)$.\\
		\setlength{\parindent}{-1.5em}
		\textbf{(6.4)} In ER model, estimated the risk loading parameter $\hat{\varphi}_{\text{ER}}^{(t)}$ by using \eqref{eq: EPP}  and \eqref{eq: allocation}, and
		$\hat{z}_{s4}^{(t)}=\hat{\mu}_s^{(t)} + \hat{\varphi}_{\text{ER}}^{(t)}\left[v_{Y_i^{(t)}}(0.95|\bm{x}_s^{(t)}) - \hat{\mu}_s^{(t)} \right].$\\
		%			\quad\quad  6.5) The risk premium prediction in each tariff class $\hat{z}_{sj}^{(t)}$ for $s=1,...,S$ and $j=1,...,J$ can be estimated \\
		%		\quad\quad \quad 	in the similar way. \\
		\setlength{\parindent}{-2em}
		\textbf{End for}.\\
		\setlength{\parindent}{-2em}
		\textbf{Return} $S^2_{sj}$, $\widehat{\text{MSE}}_{sj}$ and $\overline{\text{MSE}}_j$ for $s=1,...,S$ and $j=1,...,J$ using (\ref{eq: barmse}), (\ref{eq: hatmse}) and (\ref{eq: samplevar}).\\
		\bottomrule
	\end{tabular}
	
	\label{tab: simulation-0}
	%\begin{tablenotes}
	%	\footnotesize
	%	\item Notes: The smallest MSE is in boldface.
	%\end{tablenotes}
	%}	
\end{table}

The simulation procedures are summarized in Table \ref{tab: simulation-0}.
Table \ref{tab: simulation-1} reports the MSE and the sample variance for the QR, PQR, QRII and ER models with respect to the values of  $\varphi$.
The results are obtained by averaging out $T=5,000$ data sets.
%\textcolor[rgb]{1.00,0.00,0.00}{从表2可以看出，相比于已有模型QR, PQR, QRII，本文提出的ER模型在不同tariff classes下的bias结果与前三种模型基本保持一致，验证了ER模型的合理性�� �表4较为详细地给出了四种模型在不同tariff classes下的MSE及其方差，进一步阐述了ER模型的优越性。具体而言：}	
%As can be seen from Table \ref{tab: simulation-1},
%In Table \ref{tab: simulation-1}, the bias results of the proposed ER model under different tariff classes are basically consistent with the first three models,
%
%
%compared with the existing models QR, PQR, QRII, the bias results of the proposed ER model under different tariff classes are basically consistent with the first three models,
%which verifies the rationality of the ER model.
%Table 4 shows the MSE and variance of the four models in different tariff classes in more detail, and further illustrates the superiority of the ER model.
%In particular:
% Overall,
{In Table \ref{tab: simulation-1},
we observe that the MSE of the ER model is lower than other three models in all cases and this improvement becomes more pronounced for higher values of $\varphi$.
 The sample variance for the ER model is smaller than other three models, indicting the estimates of the risk premium are stable with
% tend to have
 the lower range of the confidence interval.}
% \textcolor[rgb]{1.00,0.00,0.00}{and the results are not stable}.
 To investigate the predictive performance among the tariff classes, Table \ref{tab: simulation-2}  reports the MSE and the sample variance for 8 tariff classes
when the true risk loading is set to be low, medium and high, e.g., $\varphi=0.05,0.10,0.15$.
 It can be seen that in all three cases, ER shows the smallest MSE for predicting risk premium in all tariff classes expect the class with the lowest risk (Class\_000) and the class with the highest risk (Class\_111).

	\begin{table}[!h]
	\small
	\centering
	\renewcommand\arraystretch{1.3}
	\caption{MSE and sample variance of the estimated risk premium with respect to the values of $\varphi$.
%		The results are obtained by averaging out $T=5000$ samples.	
	}
	\setlength{\tabcolsep}{4mm}{
		% Table generated by Excel2LaTeX from sheet 'mse.all'
\begin{tabular}{ccccccccc}
	\toprule
		\multirow{2}[1]*{{\bf $\varphi$}} & \multicolumn{ 2}{c}{{\bf QR}} & \multicolumn{ 2}{c}{{\bf PQR}} & \multicolumn{ 2}{c}{{\bf QRII}} & \multicolumn{ 2}{c}{{\bf ER}} \\
	\cline{2-9}
	\multicolumn{ 1}{c}{{\bf }} &  {\bf $\overline{\text{MSE}}_1$} & {\bf $S_{\overline{\text{MSE}}_1}^2$} &  {\bf $\overline{\text{MSE}}_2$} & {\bf $S_{\overline{\text{MSE}}_2}^2$} & {\bf $\overline{\text{MSE}}_3$} & {\bf $S_{\overline{\text{MSE}}_3}^2$} &  {\bf $\overline{\text{MSE}}_4$} & {\bf $S_{\overline{\text{MSE}}_4}^2$} \\
	\hline
0.02  & 361.51 & 36.16 & 360.91 & 36.03 & 435.41 & 72.75 & \textbf{358.29} & \textbf{35.49 }\\
0.04  & 373.78 & 43.27 & 372.54 & 43.07 & 429.03 & 69.21 & \textbf{367.57} & \textbf{41.97} \\
0.06  & 379.10 & 42.74 & 377.62 & 42.63 & 447.44 & 81.84 & \textbf{370.32} & \textbf{41.14} \\
0.08  & 392.06 & 45.58 & 388.98 & 45.01 & 476.18 & 89.61 & \textbf{377.90} & \textbf{41.80} \\
0.10   & 403.41 & 47.01 & 398.64 & 46.32 & 473.18 & 87.49 & \textbf{384.42} & \textbf{43.38 }\\
0.12  & 427.02 & 54.48 & 421.43 & 53.31 & 498.18 & 89.84 & \textbf{405.02} & \textbf{49.11} \\
0.14  & 444.63 & 63.44 & 436.40 & 61.70 & 507.09 & 104.13 & \textbf{412.92} & \textbf{54.30} \\
	\bottomrule
\end{tabular}

		\label{tab: simulation-1}
		\begin{tablenotes}
			\footnotesize
			\item Notes: The smallest value is in boldface.
%			Best performance is in boldface.
		\end{tablenotes}
	}	
\end{table}

	\begin{table}[!h]
	\small
	\centering
	\renewcommand\arraystretch{1.3}
	\caption{MSE and sample variance of the estimated risk premium for 8 tariff classes in the case of $\varphi=0.05,0.10,0.15$.
}
	\setlength{\tabcolsep}{2mm}{
% Table generated by Excel2LaTeX from sheet 'mse.all'
\begin{tabular}{cccccccccc}
	\toprule
	\multirow{2}[1]*{{\bf Cases}} & 	\multirow{2}[1]*{{\bf Traif Classes}} & \multicolumn{ 2}{c}{{\bf QR}} & \multicolumn{ 2}{c}{{\bf PQR}} & \multicolumn{ 2}{c}{{\bf QRII}} & \multicolumn{ 2}{c}{{\bf ER}} \\
	\cline{3-10}	
	\multicolumn{ 1}{c}{{\bf }} & \multicolumn{ 1}{c}{{\bf }} &        $\widehat{\text{MSE}}_{s1}$ &        $S_{\hat{z}_{s1}}^2$ &        $\widehat{\text{MSE}}_{s2}$ &        $S_{\hat{z}_{s2}}^2$ &        $\widehat{\text{MSE}}_{s3}$ &        $S_{\hat{z}_{s3}}^2$ &        $\widehat{\text{MSE}}_{s4}$ &        $S_{\hat{z}_{s4}}^2$ \\
\hline
	\multirow{8}[1]*{\tabincell{c}{Low risk loading \\$\varphi=0.05$}}           & \multicolumn{1}{c}{Class\_000} & \textbf{122.29} & 120.64 & 122.90 & 121.65 & 197.79 & 197.66 & 129.29 & 127.36 \\
& Class\_001 & 312.93 & 311.73 & 312.61 & 311.83 & 480.75 & 480.44 & \textbf{309.81} & 309.78 \\
& Class\_010 & 151.48 & 151.24 & 150.96 & 150.89 & 230.59 & 230.45 & \textbf{150.08} & 149.95 \\
& Class\_011 & 401.49 & 401.49 & 398.00 & 397.99 & 486.41 & 486.40 & \textbf{388.02} & 387.60 \\
& Class\_100 & 298.48 & 297.67 & 297.42 & 296.96 & 447.28 & 446.52 & \textbf{293.37} & 293.32 \\
& Class\_101 & 751.21 & 750.64 & 745.01 & 744.51 & 1022.79 & 1022.18 & \textbf{715.89} & 715.87 \\
& Class\_110  & 294.80 & 294.77 & 292.88 & 292.82 & 331.60 & 331.39 & \textbf{287.23} & 286.78 \\
& Class\_111 & 713.38 & 713.38 & 712.08 & 711.91 & \textbf{451.70} & 450.13 & 704.46 & 699.54 \\
\hline
	\multirow{8}[1]*{\tabincell{c}{Medium  risk loading \\$\varphi=0.10$}}  & \multicolumn{1}{c}{Class\_000} & 153.25 & 149.92 & \textbf{153.22} & 150.95 & 217.13 & 217.07 & 169.93 & 165.53 \\
& Class\_001 & 374.04 & 372.70 & 371.76 & 371.25 & 521.01 & 520.66 & \textbf{367.09} & 365.39 \\
& Class\_010 & 178.00 & 176.75 & 176.06 & 175.64 & 242.34 & 242.31 & \textbf{173.14} & 172.74 \\
& Class\_011 & 426.94 & 426.94 & 420.54 & 420.52 & 483.85 & 483.85 & \textbf{400.45} & 398.77 \\
& Class\_100 & 329.62 & 328.47 & 327.83 & 327.39 & 449.54 & 449.15 & \textbf{323.33} & 322.18 \\
& Class\_101 & 811.47 & 811.46 & 802.39 & 802.39 & 1035.35 & 1034.75 & \textbf{746.70} & 744.05 \\
& Class\_110  & 320.93 & 320.91 & 316.72 & 316.62 & 355.94 & 355.82 & \textbf{301.91} & 300.58 \\
& Class\_111 & 729.27 & 728.19 & 721.87 & 721.82 & \textbf{494.19} & 493.46 & 703.28 & 691.95 \\
\hline
	\multirow{8}[1]*{\tabincell{c}{High risk loading \\$\varphi=0.15$} } & \multicolumn{1}{c}{Class\_000} & \textbf{164.02} & 156.47 & 165.21 & 160.12 & 241.37 & 241.30 & 193.22 & 183.59 \\
& Class\_001 & 395.15 & 392.43 & 393.90 & 392.91 & 562.78 & 562.42 & \textbf{386.94} & 382.35 \\
& Class\_010 & 202.52 & 199.44 & 197.83 & 196.75 & 274.43 & 274.35 & \textbf{194.58} & 193.96 \\
& Class\_011 & 458.26 & 458.24 & 448.35 & 448.25 & 532.89 & 532.87 & \textbf{415.47} & 411.33 \\
& Class\_100 & 352.95 & 350.06 & 349.32 & 348.25 & 490.51 & 490.14 & \textbf{337.49} & 334.68 \\
& Class\_101 & 884.55 & 884.45 & 868.81 & 868.58 & 1163.47 & 1162.73 & \textbf{783.64} & 774.65 \\
& Class\_110  & 327.06 & 327.06 & 320.44 & 320.36 & 358.63 & 358.57 & \textbf{301.07} & 298.66 \\
& Class\_111 & 766.33 & 760.70 & 759.63 & 758.33 & \textbf{538.33} & {537.46} & 734.76 & 716.74 \\
\bottomrule	
\end{tabular}
		\label{tab: simulation-2}
				\begin{tablenotes}
			\footnotesize
			\item Notes: Class\_000 represents the tariff class for $x_1=0,x_2=0,x_3=0$ with the smallest pure premium, and Class\_111
			\item represents
			  represents $x_1=1,x_2=1,x_3=1$ with the largest pure premium.
		\end{tablenotes}
	}	
\end{table}

%\begin{figure}[!h]
%	\centering
%	\includegraphics[scale = 0.65]{fig-bias-caseI.pdf}
%	\caption{The boxplot of the bias of the estimated risk premium for 8 tariff classes from $T=5,000$ simulated samples in the Case 1.
%}
%	\label{fig-bias-caseI}
%\end{figure}
%
%
%\begin{figure}[!h]
%	\centering
%	\includegraphics[scale = 0.65]{fig-bias-caseII.pdf}
%	\caption{The boxplot of the bias of the estimated risk premium for 8 tariff classes from $T=5,000$ simulated samples in the case II.
%	}
%	\label{fig-bias-caseII}
%\end{figure}

\clearpage
\section{Case study: Australian automobile insurance}\label{sec: results}
%{
	This section will explore the plausibility of the ER model by providing deeper insights into the model performance comparison.
%	comparing the results from several existing studies. Deeper insights are provided into the model performance comparison.
	
%	first explore the plausibility of the ER model by comparing the results from several existing studies. Next, deeper insights are provided into the model performance comparison.
	
\subsection{Data description}
This study's data set contains information on full comprehensive Australian automobile insurance policies between years 2004 and 2005, which comes from \citet{de2008generalized}; the same data set is analyzed in  \citet{heras2018application} and \citet{baione2019, baione2020application}.
The insurance portfolio contains 67,856 policies, of which 4,624 have at least one claim. Each claim record consists of an aggregate claim amount (Claimcst0), claim numbers (Numclaims), occurrence of claim (Clm), risk exposure, and several covariates, such as age of policyholder, age of vehicle, value of vehicle, area of residence, and body type of vehicle.
For simplification and comparative purposes, we consider the same covariates as \citet{heras2018application} in the following application: age of vehicle (Veh\_age) and age of driver (Agecat).

\begin{figure}[htb]
	\centering
	\includegraphics[scale = 0.55]{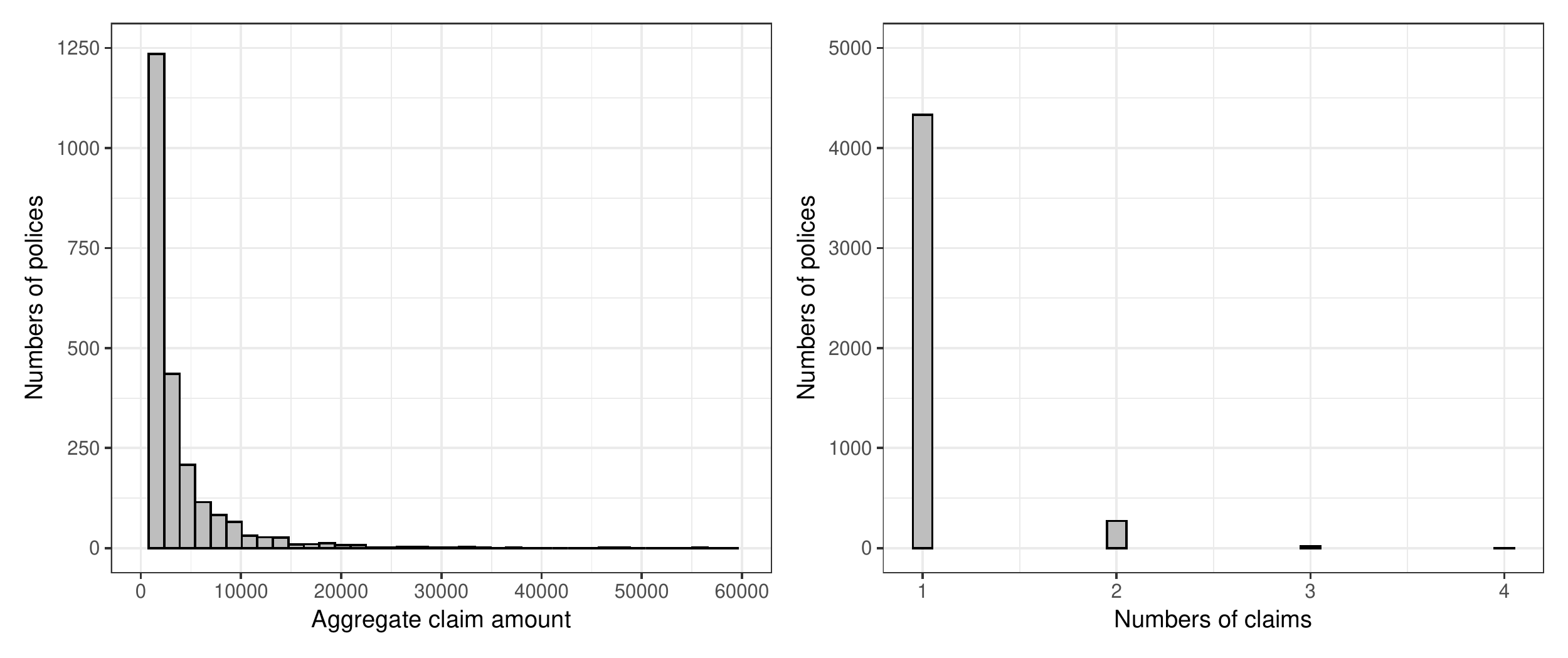}
	\caption{Histogram of the (positive) aggregate claim amount of individual polices (\textit{left panel})  and bar-plot of the claim numbers for those policies that have one or more claims (\textit{right panel}) .}
	%		Predictive distribution of aggregate claim amount (left panel) and QQ-plot of aggregate claim amount (right panel) of the portfolio.}
	\label{fig1}
\end{figure}

{The variables in the data set are listed in Table \ref{tab 2}. For each policy, we define the aggregate claim amount as the sum of the cost of all claims submitted by each policyholder, assuming that the aggregate amount is zero if the policy has no claim. A histogram of the (positive) aggregate claim amount is given in the left panel of Figure \ref{fig1}.
	%For clarity, the horizontal axis is truncated at \$50,000.
	%Only one claim above
	%A total of 65 claims between \$15,000 and \$57,000 are omitted from this display.
	A bar-plot of the claim numbers for those policies that have one or more claims is given in right panel of Figure \ref{fig1}. In this portfolio, most of the policies, up to 93.19\%, have only one claim each and only 0.002947\% have four claims each.
}
\begin{table}[htbp]
	\small
	\centering
	\caption{Description of Variables.}
	\renewcommand\arraystretch{1.5}
	\begin{tabular*}{\hsize}{@{}@{\extracolsep{\fill}}lll@{}}
		\toprule
		\textbf{Variables} & \textbf{Type} & \textbf{Description}\\
		\hline
		Agecat&	Categorical&	Driver's age category: 1 (youngest), 2, 3, 4, 5, 6\\
		Veh\_age	&Categorical&	Age of vehicle: 1 (youngest), 2, 3, 4\\
		Exposure&	Continuous&	Policy years (between 0 and 1)\\
		Clm	&Discrete&	Occurrence of claim (0 = no, 1 = yes)\\
%		Numclaims&	Discrete&	Numbers of claims(0, 1, 2, 3,$\cdots$)\\
		Claimcst0&	Continuous	&Aggregate claim amount of a policy (0 if no claim)\\  \bottomrule
	\end{tabular*}
	\label{tab 2}
\end{table}

%\newpage
\subsection{Calculation of the individual risk premium}

The first step of the proposed method concerns the estimate of the pure premium.
This can be easily performed by a two-part GLMs assuming a binomial distribution with a logistic link function for the claim probability that is corrected for risk exposure, and a Gamma distribution with a log link function for the non-zero claim amount.
Table \ref{tab: estimates-glms} shows the parameter estimates and standard errors for the two-part GLMs with the two most important risk factors, Veh\_age and Agecat, and with four and six levels \footnote{We only take into account these two risk factors because it simplifies the analysis. Further, most of the levels are significantly different from zero, and the same risk factors are used in
	\citet{heras2018application}.
} respectively.
%The actual risk exposure is considered in the model.
For the logistic regression part, all the parameters are highly significant (i.e. p-values less than 0.05), except for the first level of Veh\_age and the sixth level of Agecat. Although most of risk factor levels are significant in the logistic regression, some of them are not significant in modelling the non-zero claim amount.

\begin{figure}[htb]
	\centering
	\includegraphics[scale = 0.65]{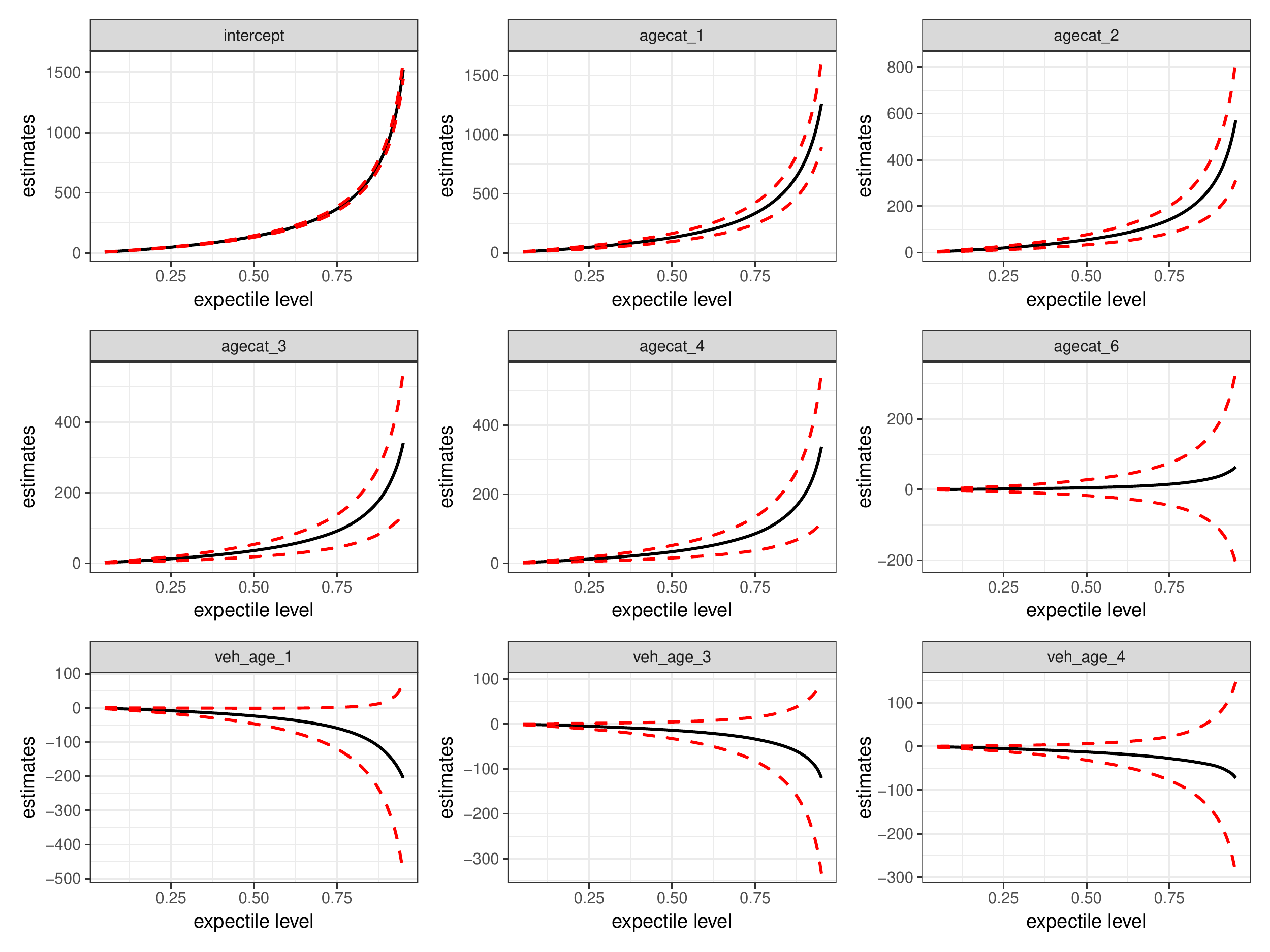}
	\caption{The estimates of parameters in ER model for expectile level varying from $\tau=0.05$ to $\tau=0.95$.
		The 95\% confidence interval are provided by the red dashed curves.
	.}
	\label{fig-ER}
\end{figure}

The second step concerns the estimate of the risk loading. It can be performed by an expectile regression. Table \ref{tab: estimates-er} reports the parameter estimates and standard errors for expectile regression where a 95\%-level expectile is considered. While all the levels of Veh\_age are not significant in modelling the 95\%-level expectile, all the levels of Agecat are highly significant (i.e., p-values less than 0.05), except  for the sixth level of Agecat.
Figure \ref{fig-ER} shows the trend of the parameters in expectile regression over the entire range of expectile levels from 0.05 to 0.95. In each graph, the central continuous line represents the parameter estimates, the external dashed lines represent the upper and lower confidence interval at level 95\% estimated by the ER. The linear combination with the intercept parameter allows the estimation of the conditional expectile of the reference tariff: agecat\_5 and veg\_age\_2. In Figure \ref{fig-ER}, we can observe that intercept and agecat parameters increase as the expectile level increases, while veh\_age parameters decrease as the expectile increases. In particular, the veh\_age parameters show a slight decrease until 0.75 and a slop afterwards.

\begin{table}[htbp]
	\small
	\centering
	\begin{threeparttable}
		\renewcommand\arraystretch{1.3}
		\caption{Parameter estimates of two-part GLMs for modelling claim probability and non-zero claim amount.}
						\setlength{\tabcolsep}{6mm}{
			\begin{tabular}{l|l|l|l|l|l}
%		\begin{tabular*}{\hsize}{@{}@{\extracolsep{\fill}}l|l|l|l|l|l|l@{}}
			\toprule
%			\textbf{Response variable}	
			\textbf{Models}	&  \textbf{Parameters}    & \textbf{Estimates}&\textbf{StdError} &\textbf{Z-value}
			& \textbf{P-value}   \\
			\hline
%			\multirow{9}*{\tabincell{l}{ occurrence of claim \\(0 = no, 1 = yes)}}		
			\multirow{9}*{\tabincell{l}{ Logistic regression}}
			 &(Intercept)    & -1.91   & 0.05    & -36.86   & 0.00 \\
			& veh\_age\_1   & -0.03   & 0.05    & -0.62    & 0.54 \\
			& veh\_age\_3   & -0.13   & 0.04    & -2.89    & 0.00 \\
			& veh\_age\_4   & -0.22   & 0.05    & -4.88    & 0.00 \\
			& agecat\_1     & 0.53    & 0.07    & 7.80     & 0.00 \\
			& agecat\_2     & 0.33    & 0.06    & 5.81     & 0.00 \\
			& agecat\_3     & 0.27    & 0.06    & 4.92     & 0.00 \\
			& agecat\_4     & 0.23    & 0.06    & 4.15     & 0.00 \\
			& agecat\_6     & 0.00    & 0.07    & -0.04    & 0.97 \\
			\hline
			\cline{2-6}
%			\multirow{9}*{\tabincell{l}{ Non-zero\\claim amount}}				
			 \multirow{9}*{\tabincell{l}{ GA regression}}  	
		&	(Intercept) &        7.46 &      0.08 &      89.33 &       0.00 \\			
			& 	veh\_age\_1 &    -0.14 &      0.08 &      -1.90 &       0.06 \\			
			& 	veh\_age\_3 &     0.11 &      0.07 &       1.53 &       0.13 \\			
			& 	veh\_age\_4 &     0.25 &      0.07 &       3.30 &       0.00 \\			
			& 	agecat\_1 &       0.57 &      0.12 &       4.94 &       0.00 \\			
			& 	agecat\_2 &       0.18 &      0.09 &       1.89 &       0.06 \\			
			& 	agecat\_3 &       0.14 &      0.09 &       1.53 &       0.13 \\			
			& 	agecat\_4 &       0.09 &      0.09 &       0.98 &       0.33 \\			
			& 	agecat\_6 &       0.04 &      0.11 &       0.38 &       0.70 \\
			\bottomrule
		\end{tabular}
	}
		\label{tab: estimates-glms}
		\begin{tablenotes}
			\footnotesize
			\item Notes: The log link function is used in the GA regression model.
%			This table in PQR reports the \textit{ilm} estimates of $\bm{\hat{\gamma} }( \tau| \bm{\hat{\theta} })$ and the asymptotic standard errors based on model \ref{eq: PQR}. For example, the estimated quantile regression coefficient with the indicator of veh\_age1 at the quantile level $\tau=74.98\%$  is $\hat{\gamma}_2(\tau)=\hat{\theta}_{21}+\hat{\theta}_{22}\tau + \hat{\theta}_{23}\tau^2=-0.14$, where $\hat{\theta}_{2j},j=1,2,3,$ denotes the estimated parameter for veh\_age1.
		\end{tablenotes}
	\end{threeparttable}
\end{table}

	\begin{table}[htbp]
	\small
	\centering
		\renewcommand\arraystretch{1.3}
		\caption{Parameter estimates of expectile regression model for modelling the aggregate claim amount data ($\tau=0.95$).}
				\setlength{\tabcolsep}{9mm}{
\begin{tabular}{l|l|l|l|l|l}
			\toprule
%		 \textbf{Response variable}   &			
		 \textbf{Parameters}    & \textbf{Estimates}&\textbf{StdError} &\textbf{Z-value}
			& \textbf{P-value}   \\
			\hline
%\multirow{9}*{\tabincell{l}{ Aggregate\\ claim amount }}			
	 (Intercept) &    1521.26 &      47.34 &      32.14 &       0.00 \\
			
veh\_age\_1 &    -205.14 &     170.58 &      -1.20 &       0.23 \\
			
			veh\_age\_3&    -120.43 &     129.99 &      -0.93 &       0.35 \\
			
			veh\_age\_4 &     -72.35 &     133.56 &      -0.54 &       0.59 \\
			
			agecat\_1 &    1260.92 &     223.62 &       5.64 &       0.00 \\
			
			agecat\_2 &     570.65 &     157.55 &       3.62 &       0.00 \\
			
			agecat\_3 &     341.32 &     123.95 &       2.75 &       0.01 \\
			
			agecat\_4 &     337.29 &     133.25 &       2.53 &       0.01 \\
			
			agecat\_6 &      63.69 &     164.83 &       0.39 &       0.70 \\
			\bottomrule
		\end{tabular}
			\label{tab: estimates-er}
%		\begin{tablenotes}
%			\footnotesize
%			\item Notes: This table in PQR reports the \textit{ilm} estimates of $\bm{\hat{\gamma} }( \tau| \bm{\hat{\theta} })$ and the asymptotic standard errors based on model \ref{eq: PQR}. For example, the estimated quantile regression coefficient with the indicator of veh\_age1 at the quantile level $\tau=74.98\%$  is $\hat{\gamma}_2(\tau)=\hat{\theta}_{21}+\hat{\theta}_{22}\tau + \hat{\theta}_{23}\tau^2=-0.14$, where $\hat{\theta}_{2j},j=1,2,3,$ denotes the estimated parameter for veh\_age1.
%		\end{tablenotes}
}

\end{table}

Table \ref{tab: prediction-riskpremium} and Table \ref{tab: prediction-riskloading} present the risk premiums and the risk loadings for 24 tariff classes predicted using the ER model, along with the GLMs, QR reported in \citet{heras2018application}, PQR reported in \citet{baione2020application} and QRII reported in \citet{baione2019}.
Five premium principles (i.e., EVPP, SDPP, QPP, TSQPP, EPP) are used to calculate the individual policy risk premiums in existing studies. For ease of comparison with the results of the other studies, we use the same estimation result of quantile regression models (i.e., QR and PQR) as given in \cite{heras2018application} and \cite{baione2020application} when the quantile level $\tau=95\%$
\footnote{
	\citet{heras2018application} applies the two-part QR model for modelling the claim probability and non-zero claim amount, and
	\citet{baione2020application} applies the two-part PQR model when the quantile level $\tau=95\%$.
	We choose the class of the Shifted Legendre polynomial (SLP) of degree 3 for ${{\gamma }_{j}}( \tau|\bm{\theta})$ in PQR model.
}.
The same total risk premium of the whole portfolio that the insurer should charge $C=\$22,206,147$ is considered in the following.

\begin{figure}[!ht]
	\centering
	\includegraphics[scale = 0.6]{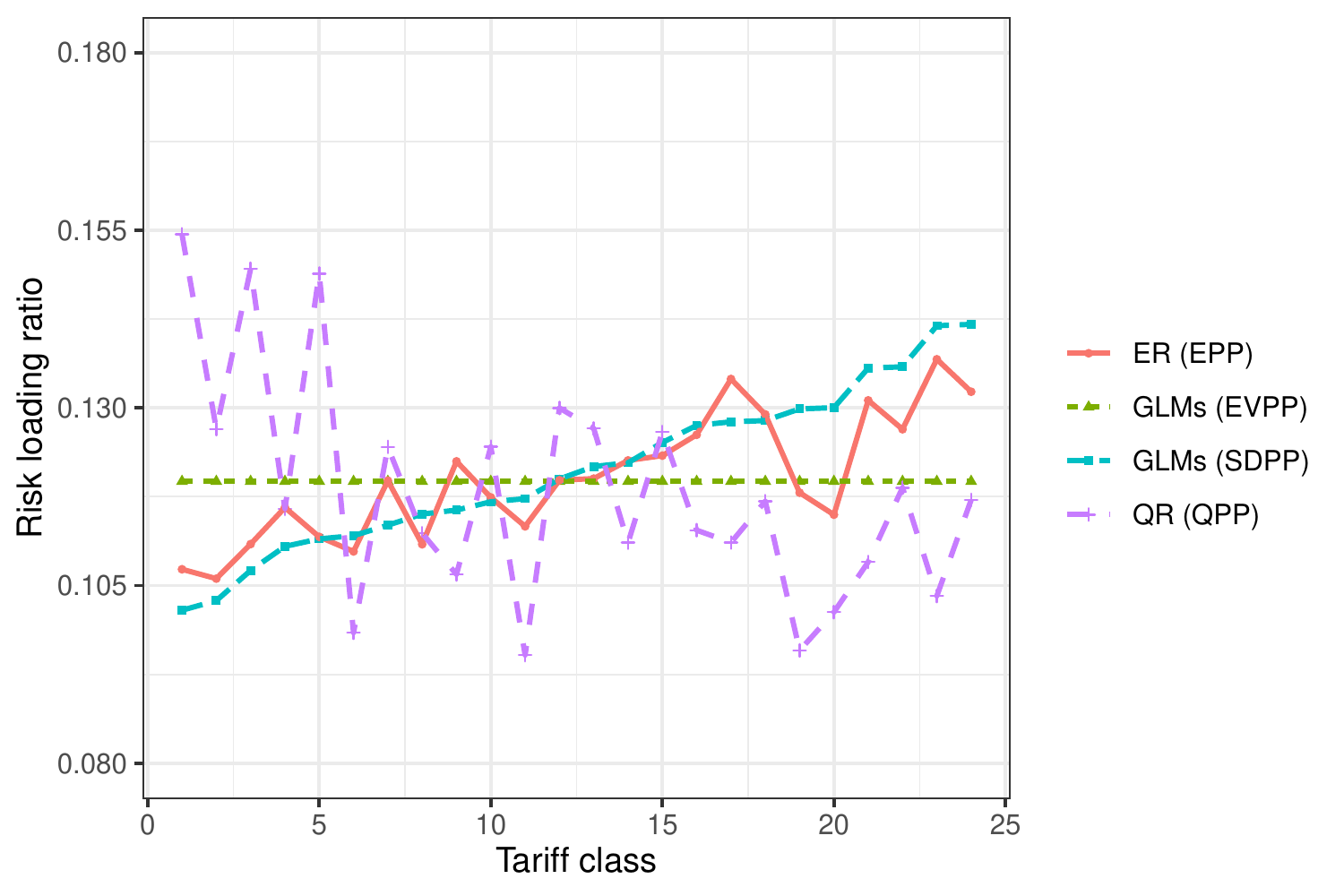}
	\caption{Risk loading ratio estimation based on the ER model, along with the results of existing studies.}
	\label{fig: comparsion}
\end{figure}

For the given $C=\$22,206,147$,
it allows us to estimate the risk loading parameter $\varphi$ in the EVPP, SDPP, QPP and EPP
using \eqref{eq: allocation}.
The result of QRII is reported for the risk loading parameter $\tau=79.08\%$ in \citet{baione2019}.
We also observe that the risk loading in GLMs is expressed as 11.97\% of the pure premium using the EVPP, and 2.31\% of the deviation using the SDPP,
as well as 3.00\% and 3.09\% of the distance between VaR and pure premium in QR and PQR using QPP, respectively. Although the PQR describes the regression coefficients' functional form parametrically depending on the order of the quantile level, QR and PQR yield similar estimates of the risk loading parameter around 3.00\%, whereas the estimate of risk loading parameter is 2.85\% in the ER model.
A possible explanation is that the expectile is considered as the basis of risk loading in EPP, which yields larger prediction than the quantile as it
gives more information about both the lower and upper tails of the distribution.
From the results we have obtained in Table \ref{tab: prediction-riskpremium} and
Table \ref{tab: prediction-riskloading}, we conclude:
\begin{enumerate}
	\item[(1)] Although two different quantile regressions (i.e. QR and PQR) are employed, the pattern of these two results is similar due to use of the same premium principle.
	
	\item[(2)]  Quantile regressions (i.e. QR and PQR) overestimate the risk loadings for high-risk classes (i.e, class 1, 2 and 3), and underestimate them for low-risk classes (i.e. class 19, 21, 23), 
	while it does not happen in the ER and GLMs.
%	both yield similar results for these classes.
%	This can be also confirmed in Figure \ref{fig: comparsion}.
%	higher 
%	risk loadings for low-risk classes (i.e. class 19, 21, 23) and lower loadings for high-risk classes (i.e, class 1, 2 and 3).
%	can result in higher risk loadings for high-risk classes and lower risk loadings for low-risk classes when the total risk is controlled, 
%	conversely the ER and GLMs yield higher risk loadings for low-risk classes (i.e. class 19, 21, 23) and lower loadings for high-risk classes (i.e, class 1, 2 and 3).
	
	\item[(3)] Larger predictions are observed in QRII. Moreover, a negative value of risk loading is also observed in class 20.
	The presence of negative risk loading might be a drawback when the TSQPP is used for classification ratemaking.
\end{enumerate}
For a graphical comparison that confirms the results in Table \ref{tab: prediction-riskpremium}, we show the predictive risk loading ratios for the 24 tariff classes based on the Figure \ref{fig: comparsion}'s competing models.
The risk loading ratio is defined as the risk loading divided by the predictive pure premium.
The risk loading ratio in GLMs (EVPP) is constant among tariff classes.
%The risk loading ratio in GLMs (EVPP) is constant among tariff classes, 
While it shows the similar pattern in ER and GLMs (SDPP),
more variation among tariff classes in ER is observed than in GLMs (SDPP), which indicates the proposed model has the
advantage of better differentiating the heterogeneity among tariff classes.
QR model has a poor model performance in sense that a opposite trend is observed.

%Meanwhile, it yields less variation among tariff classes in ER than in QR model.

%Thus, the ER model occupies an interesting position between QR and GLMs.

%More variation among tariff classes in ER model is observed  than GLMs (SDPP) .

%While it shows 
%A similar trend can be observed both in the ER model and GLMs (SDPP), 
%opposite trend

%While 

%More specially, More variation among tariff classes in ER model is observed  than GLMs (SDPP) .

%in the risk loading ratios among tariff classes obtained by ER model is observed  than GLMs (SDPP) .

%In addition,
%higher risk loading ratio in high-risk class is observed in QR, whereas lower risk loading ratio in high-risk class is observed in ER.
%More variation in the risk loading ratios among tariff classes is observed QRII (\textit{right panel} in Figure \ref{fig: comparsion}) than other five models (\textit{left panel} in Figure \ref{fig: comparsion}).
%The risk loading ratio in GLMs (EVPP) remains to be constant.
%Meanwhile, it yields less variation among tariff classes in GLMs (SDPP) than ER model.

\begin{table}[p]
	\small
	\begin{center}
		\caption{Classification risk premiums predicted by different models.}
		\renewcommand\arraystretch{1.3}
				\setlength{\tabcolsep}{2.5mm}{
%		\begin{tabular*}{\hsize}{@{}@{\extracolsep{\fill}}l|l|l|l|l|l|l|l|l|l@{}}
\begin{tabular}{lllllllll}
			\toprule
	\multirow{2}[1]*{\tabincell{l}{\textbf{Tariff}\\ \textbf{Class}}}
	& \multicolumn{1}{c}{\multirow{2}[0]{*}{$\hat{p}_i$}} & \multicolumn{1}{c}{\multirow{2}[0]{*}{\tabincell{l}{\textbf{Pure}\\ \textbf{premium}}}}
	& \multicolumn{1}{l}{\textbf{ER}} & \multicolumn{2}{c}{\textbf{GLMs}} & \multicolumn{1}{l}{\textbf{QR}} & \multicolumn{1}{l}{\textbf{PQR}} & \multicolumn{1}{l}{\textbf{QRII}} \\
	\cline{4-9}
	&       &       & {\tabincell{l}{\textbf{EPP}\\$\varphi=2.85\%$}}
& \tabincell{l}{\textbf{EVPP}\\$\varphi=11.97\%$}&\tabincell{l}{\textbf{SDPP}\\$\varphi=2.31\%$}
	& \tabincell{l}{{\textbf{QPP}}\\$\varphi = 3.00\%$}
	& \tabincell{l}{\textbf{QPP} \\ $\varphi =  3.09\%$}
	& \tabincell{l}{\textbf{TSQPP} \\ $\tau =  79.08\%$}\\
	\hline
    1-V2A1 & 0.798 & 522.88 & 578.98 & 585.45 & 575.97 & 603.63 & 602.93 & 728.58 \\
2-V1A1 & 0.803 & 484.58 & 535.93 & 542.56 & 534.43 & 546.13 & 550.16 & 585.84 \\
3-V3A1 & 0.818 & 484.98 & 538.73 & 543.01 & 536.93 & 557.52 & 562.59 & 771.13 \\
4-V2A2 & 0.828 & 355.42 & 396.64 & 397.95 & 394.69 & 396.57 & 396.55 & 415.44 \\
5-V4A1 & 0.831 & 491.20 & 546.14 & 549.98 & 546.00 & 564.34 & 570.14 & 784.62 \\
6-V1A2 & 0.833 & 329.07 & 365.21 & 368.45 & 365.94 & 361.44 & 362.93 & 333.73 \\
7-V2A3 & 0.837 & 302.31 & 338.52 & 338.49 & 336.62 & 339.95 & 339.79 & 381.75 \\
8-V1A3 & 0.841 & 279.83 & 310.84 & 313.31 & 312.03 & 311.27 & 310.90 & 306.58 \\
9-V2A4 & 0.843 & 296.12 & 332.39 & 331.56 & 330.36 & 327.68 & 329.20 & 346.93 \\
10-V3A2 & 0.846 & 328.44 & 367.01 & 367.75 & 366.80 & 369.35 & 367.72 & 438.08 \\
11-V1A4 & 0.847 & 274.05 & 305.11 & 306.84 & 306.18 & 300.14 & 301.50 & 278.58 \\
12-V3A3 & 0.853 & 279.08 & 312.52 & 312.47 & 312.57 & 315.36 & 314.73 & 402.14 \\
13-V4A2 & 0.857 & 331.81 & 371.65 & 371.52 & 372.21 & 373.98 & 371.19 & 444.62 \\
14-V3A4 & 0.859 & 273.17 & 306.67 & 305.86 & 306.58 & 303.50 & 304.56 & 365.21 \\
15-V4A3 & 0.865 & 281.74 & 316.47 & 315.45 & 316.99 & 317.41 & 317.31 & 407.84 \\
16-V4A4 & 0.87  & 275.65 & 310.44 & 308.63 & 310.80 & 306.74 & 306.85 & 370.22 \\
17-V2A5 & 0.871 & 215.94 & 244.89 & 241.78 & 243.59 & 239.92 & 238.77 & 273.48 \\
18-V2A6 & 0.871 & 234.28 & 264.52 & 262.31 & 264.32 & 261.66 & 259.32 & 257.69 \\
19-V1A5 & 0.874 & 199.67 & 223.25 & 223.57 & 225.60 & 218.81 & 219.07 & 219.41 \\
20-V1A6 & 0.875 & 216.63 & 241.53 & 242.55 & 244.80 & 238.56 & 238.00 & 206.73 \\
21-V3A5 & 0.884 & 198.53 & 224.55 & 222.28 & 225.45 & 220.03 & 219.79 & 286.91 \\
22-V3A6 & 0.885 & 215.38 & 242.73 & 241.15 & 244.62 & 240.96 & 239.05 & 270.33 \\
23-V4A5 & 0.894 & 199.86 & 227.21 & 223.77 & 228.15 & 220.55 & 220.20 & 290.16 \\
24-V4A6 & 0.894 & 216.82 & 245.49 & 242.76 & 247.55 & 242.19 & 239.68 & 273.39 \\
	\bottomrule
\end{tabular}%

%		\end{tabular}
		\label{tab: prediction-riskpremium}
		\begin{tablenotes}
			\footnotesize
			\item Notes: All the results are obtained for one risk exposure (i.e. one policy year). 			
			24 tariff classes have been subdivided according to the two risk factor: Veh\_age and Agecat.
			The symbol V1A1 in Column 1 denotes the policyholder with Veh\_age and Agecate both in the level 1.
			Column 2 reports the predicted probability of incurring no claim.
			Column 3 reports the predicted pure premium.
			QR and PQR denote two quantile regressions discussed in \citet{heras2018application} and \citet{baione2020application} respectively.
			QRII represents the two-stage quantile regression proposed in \citet{baione2019}.
			ER denotes the expectile regression model.
			EVPP, SDPP, QPP, EPP, and TSQPP denote various premium principles listed in Table \ref{tab: Literature}.
		\end{tablenotes}
	}
	\end{center}
\end{table}

\begin{table}[p]
	\small
	\caption{Classification risk loadings predicted by different models.}
		\renewcommand\arraystretch{1.3}
\setlength{\tabcolsep}{2.5mm}{
	%		\begin{tabular*}{\hsize}{@{}@{\extracolsep{\fill}}l|l|l|l|l|l|l|l|l|l@{}}
\begin{tabular}{lllllllll}
	\toprule
	\multirow{2}[1]*{\tabincell{l}{\textbf{Tariff}\\ \textbf{Class}}}
	& \multicolumn{1}{c}{\multirow{2}[0]{*}{$\hat{p}_i$}} & \multicolumn{1}{c}{\multirow{2}[0]{*}{\tabincell{l}{\textbf{Pure}\\ \textbf{premium}}}}
	& \multicolumn{1}{l}{\textbf{ER}} & \multicolumn{2}{c}{\textbf{GLMs}} & \multicolumn{1}{l}{\textbf{QR}} & \multicolumn{1}{l}{\textbf{PQR}} & \multicolumn{1}{l}{\textbf{QRII}} \\
	\cline{4-9}
	&       &       & {\tabincell{l}{\textbf{EPP}\\$\varphi=2.85\%$}}
	& \tabincell{l}{\textbf{EVPP}\\$\varphi=9.73\%$}&\tabincell{l}{\textbf{SDPP}\\$\varphi=2.31\%$}
	& \tabincell{l}{{\textbf{QPP}}\\$\varphi = 3.00\%$}
	& \tabincell{l}{\textbf{QPP} \\ $\varphi =  3.09\%$}
	& \tabincell{l}{\textbf{TSQPP} \\ $\tau =  79.08\%$}\\
	\hline
1-V2A1 & 0.798 & 522.88 & 56.10 & 62.57 & 53.09 & 80.75 & 80.05 & 205.70 \\
2-V1A1 & 0.803 & 484.58 & 51.35 & 57.99 & 49.85 & 61.55 & 65.58 & 101.26 \\
3-V3A1 & 0.818 & 484.98 & 53.75 & 58.03 & 51.96 & 72.54 & 77.61 & 286.15 \\
4-V2A2 & 0.828 & 355.42 & 41.22 & 42.53 & 39.28 & 41.15 & 41.13 & 60.02 \\
5-V4A1 & 0.831 & 491.20 & 54.94 & 58.78 & 54.80 & 73.14 & 78.95 & 293.42 \\
6-V1A2 & 0.833 & 329.07 & 36.13 & 39.38 & 36.86 & 32.37 & 33.86 & 4.66 \\
7-V2A3 & 0.837 & 302.31 & 36.21 & 36.18 & 34.31 & 37.64 & 37.48 & 79.43 \\
8-V1A3 & 0.841 & 279.83 & 31.01 & 33.49 & 32.20 & 31.44 & 31.07 & 26.76 \\
9-V2A4 & 0.843 & 296.12 & 36.27 & 35.43 & 34.24 & 31.56 & 33.08 & 50.81 \\
10-V3A2 & 0.846 & 328.44 & 38.56 & 39.30 & 38.36 & 40.90 & 39.27 & 109.64 \\
11-V1A4 & 0.847 & 274.05 & 31.06 & 32.79 & 32.13 & 26.09 & 27.45 & 4.53 \\
12-V3A3 & 0.853 & 279.08 & 33.44 & 33.40 & 33.49 & 36.28 & 35.65 & 123.06 \\
13-V4A2 & 0.857 & 331.81 & 39.83 & 39.71 & 40.40 & 42.17 & 39.37 & 112.80 \\
14-V3A4 & 0.859 & 273.17 & 33.49 & 32.69 & 33.40 & 30.33 & 31.39 & 92.04 \\
15-V4A3 & 0.865 & 281.74 & 34.73 & 33.71 & 35.25 & 35.67 & 35.57 & 126.11 \\
16-V4A4 & 0.870 & 275.65 & 34.79 & 32.98 & 35.16 & 31.09 & 31.21 & 94.57 \\
17-V2A5 & 0.871 & 215.94 & 28.95 & 25.84 & 27.65 & 23.98 & 22.83 & 57.54 \\
18-V2A6 & 0.871 & 234.28 & 30.24 & 28.03 & 30.04 & 27.38 & 25.05 & 23.41 \\
19-V1A5 & 0.874 & 199.67 & 23.58 & 23.89 & 25.93 & 19.14 & 19.40 & 19.73 \\
20-V1A6 & 0.875 & 216.63 & 24.91 & 25.92 & 28.17 & 21.93 & 21.38 & -9.89 \\
21-V3A5 & 0.884 & 198.53 & 26.02 & 23.76 & 26.92 & 21.50 & 21.26 & 88.38 \\
22-V3A6 & 0.885 & 215.38 & 27.35 & 25.77 & 29.25 & 25.58 & 23.67 & 54.95 \\
23-V4A5 & 0.894 & 199.86 & 27.35 & 23.92 & 28.29 & 20.69 & 20.34 & 90.31 \\
24-V4A6 & 0.894 & 216.81 & 28.68 & 25.94 & 30.74 & 25.38 & 22.86 & 56.57 \\
    		\bottomrule						
    		\end{tabular}
    	}
		\label{tab: prediction-riskloading}
\end{table}

 \subsection{Performance Comparison}
 \label{subsec: Comparison}
For model comparison, the frequently used loss functions, e.g., the root mean square error or mean square error, are not quite informative enough to capture the difference between the predictive values and the corresponding
outcomes for individual claim data, due to the high proportions of zeros and the highly right-skewed
features.
%for capturing the differences between predictions and observations, due to the high proportions of zeros and the skewed heavy-tailed distribution of the positive losses.
For this reason,
we consider the ordered Lorenz curve and the associated Gini index as a statistical measure of the association between the premium and loss distributions in non-life insurance, through which different predictive models can be compared \citep{frees2011summarizing,yang2018insurance,shi2018pair}.

The ordered Lorenz curve is a plot with the ordered loss distribution on the vertical axis and the ordered premium
distribution on the horizontal axis both sorted based on the relative premium (``competing premium''/``base premium'' ),
where ``base premium'' is calculated using an existing premium prediction model and ``competing premium'' is calculated using an alternative premium prediction.
The associated Gini index is defined as twice the area between the ordered Lorenz curve and the line of equality.
The logic behind this idea is that a score with a greater Gini index produces a greater separation among the observations.
In other words, a higher Gini index indicates greater ability to distinguish good risks from bad risks.
Following  \citet{frees2011summarizing} and \citet{yang2018insurance},
the key to selecting the most favorable model is first to specify the prediction from each model as a base premium and use the prediction from the remaining models as the competing premium.

\begin{figure}[!ht]
	\centering
	\includegraphics[scale = 0.8]{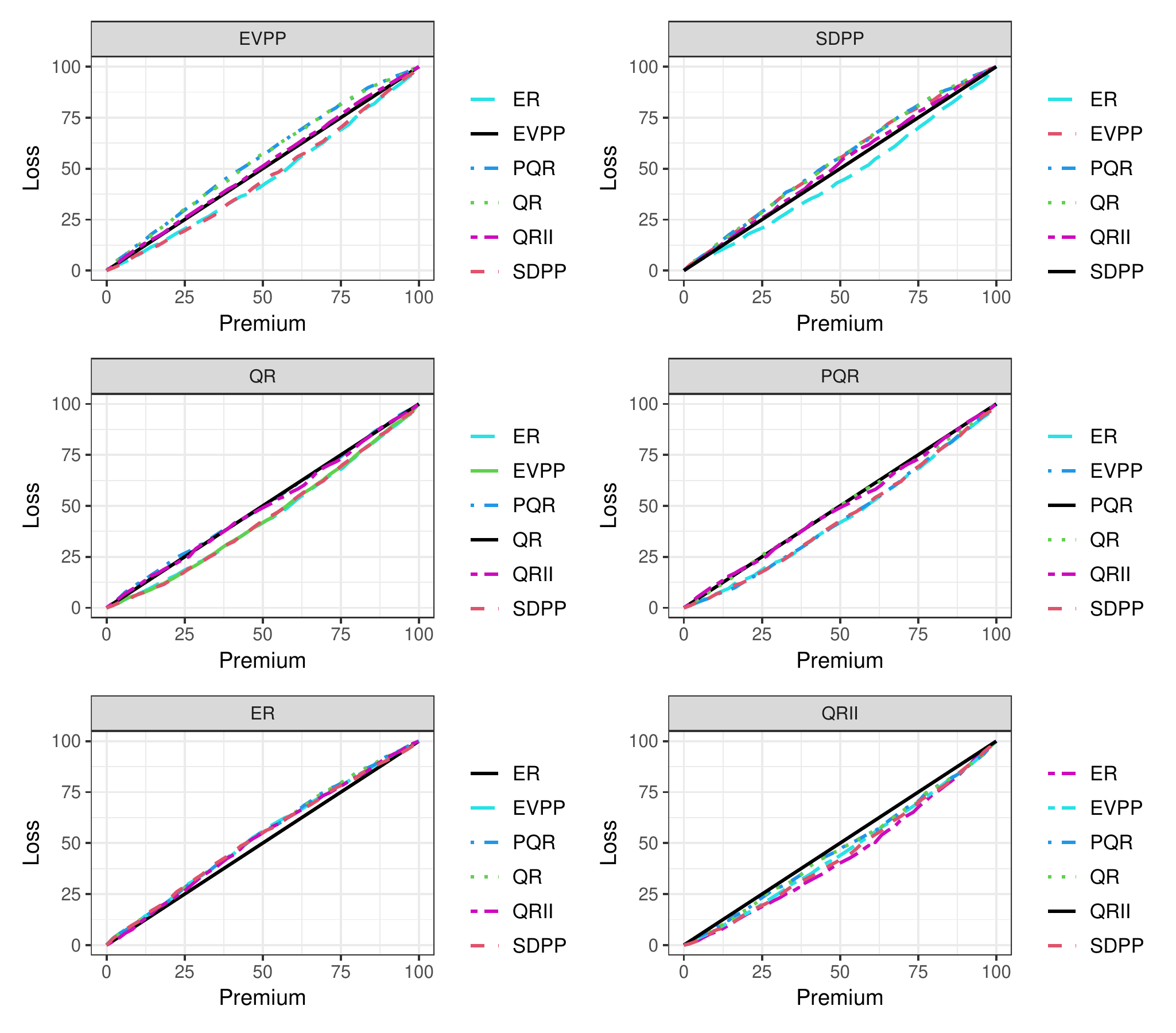}
	\caption{The ordered Lorenz curves for the automobile claim data.}
	\label{fig-gini2}
\end{figure}

We calculate the risk premium for each individual policyholder based on GLMs, QR, QRII, PQR and ER models, taking the actual exposure and risk factors of insured $i$ into account \footnote{Note that the risk premium cannot be estimated correctly in QR and PQR when the estimated probability of having no claim $\hat{p}_i=1-{{w}_{i}}\frac{\exp (\bm{x}_{i}\bm{\hat{\alpha} })}{1+\exp (\bm{x}_{i}\bm{\hat{\alpha} })}>0.95$
	if the actual risk exposure is taken into account; see subsection \ref{subsec: models} for more details.
	Thus, in order to compare all competing ratemaking methods,  we eliminate 27,173 observations from the data for prediction.
}.
In Table \ref{tab: prediction-riskpremium}, the Gini indices and standard errors are reported by averaging the results from 20 random samples splitting when specifying different combinations of base premium and competing premium.
The ``mini-max'' strategy is usually used to select the ``best'' model, that is, we first select the model that provides the largest Gini index for each base premium, and then the smallest one taking over competing premiums.
We observe that the maximal Gini index is 9.68, 8.33, 11.58, 11.64, and 0 when using the
GLMs (i.e., EVPP and SDPP), QR, PQR, ER, and QRII as the base premium.
ER with the proposed EPP has the smallest maximum Gini index at 0, hence it is most robust to the alternative model \footnote{
If taking all observations in the data into account,
we can see that ER model also shows the better model performance than QRII and GLMs (i.e., EVPP and SDPP).
}.
Figure \ref{fig-gini2} displays the ordered Lorenz curves of the proposed models.
It is obvious that when the ER model is selected as the base premium,
the area between the
line of equality and the ordered Lorentz curve is smaller than the competing models, indicating
that the ER model represents the most favorable choice again.
%{In order to eliminate of the estimate of the total risk premium of the portfolio,
%we also compare the results of all competing models when the same $C$ is considered.
%Table \ref{tab: gini2} reports the Gini indices and standard errors for all the models in the case of $C=21,977,805$.
%Figure \ref{fig-gini2} displays the ordered Lorenz curves of the competing models.
%Again, ER model with the proposed expectile premium principle has a best model performance.
%}

\begin{table}[htb!]
	\small
	\centering
	\renewcommand\arraystretch{1.5}
	\caption{The averaged Gini indices and standard errors in the
		automobile claim data example based on 20 random splits.
%		The results are reported for all the models in the case of $C=21,977,805$.
	}
	\setlength{\tabcolsep}{0.4mm}{
		\begin{tabular}{c|cccccc|c}
			\toprule
			\multirow{2}[2]{*}{{\textbf{Base} \textbf{premium}}} & \multicolumn{6}{c|}{\textbf{Competing} \textbf{premium}} & \multirow{2}[1]*{{\textbf{Mini-max}}} \\
			\cline{2-7}
& \textbf{GLMs (EVPP)} & \textbf{GLMs (SDPP)} & \textbf{QR} & \textbf{PQR} & \textbf{ER} & \textbf{QRII} & 	\\
\hline
\textbf{GLMs (EVPP)} & 0 (0) & 8.86 (1.94) & -9.64 (1.90) & -10.06 (1.92) & 9.68 (2.04) & -1.95 (1.97) &9.68 \\
\textbf{GLMs (SDPP)} & -7.65 (1.94) & 0 (0) & -8.41 (1.91) & -8.51 (1.92) & 8.33 (2.08) & -3.41 (1.93) & 8.33 \\
\textbf{QR} & 11.29 (1.90) & 11.25 (1.90) & 0 (0) & -0.78 (2.07) & 11.58 (1.97) & 0.98 (2.06) & 11.58 \\
\textbf{PQR} & 11.62 (1.92) & 11.26 (1.92) & 1.1 (2.07) & 0 (0) & 11.64 (1.98) & 1.03 (2.07) & 11.64 \\
\textbf{ER} & -6.36 (2.03) & -6.11 (2.07) & -6.65 (1.98) & -6.78 (1.98) & 0 (0) & -5.15 (1.91) & \textbf{0.00}\\
\textbf{QRII} & 8.62 (1.97) & 10.3 (1.92) & 5.66 (2.06) & 5.54 (2.07) & 12.83 (1.90) & 0 (0) & 12.83 \\
			\bottomrule
		\end{tabular}
	}
	\label{tab: gini2}
\end{table}

\newpage
\section{Conclusion }\label{sec: conclusion}
%The need for pure premium prediction for modelling insurance claims is well-recognized by actuaries; but practice tools to study risk premium prediction based on an
%arbitrary set of risk factors are very limited.
In this paper, we propose the expectile regression based on a new Expectile Premium Principle as an important complement to the conventional GLMs as well as quantile regression in the ratemaking mechanism. Our approach has the following advantages:
(1) The proposed EPP is a coherent risk measure, thus overcoming the drawbacks of QPP discussed in the recent literature.
(2) The risk premium obtained in EPP depends on the shape of the entire distribution and contains more information about both upper and lower tails of distribution.
%(3) EPP is a more conversational risk measure for the insurers and regulators as expectile is
%more sensitive to the extreme data than quantile.
(3) The proposed ratemaking method is computationally efficient. It enables the estimation of the conditional expectile of the aggregate claim amount for all expectile levels in one run, thus allowing us not to lose efficiency in the risk premium estimations.
(4) The simulation and empirical result both suggest that expectile regression outperforms the conventional GLMs and quantile regressions,
since it can better differentiate the heterogeneity among tariff classes and has a greater ability to distinguish high risks from low risks for the insurers.

It is worth mentioning that the applications of the expectile regressions can go beyond the insurance premium prediction and be of interest to researchers in many other fields including capital allocation, catastrophe risk modelling, and loss reserve assessment in actuarial science.
%Therefore,
Some extensions of this method with expectile regressions for modelling multivariate loss data
%tail risk measurement based on extreme value theory
could be studied in future research.

\section*{Acknowledgements}
The authors contributed equally to the work.
The authors acknowledge the
National Social Science Fund of China (Grant No. 16ZDA052),
the Fundamental Research Funds for the Central Universities in UIBE (Grant No. 17QD11),
and the National Natural Science Fund
of China (Grant No. 71901064).

\section*{Declaration of interest }
We declare that there is no potential conflict of interest in the paper.

%\begin{flushleft}
%	Jan Beirlant\\
%	\emph{Dept. of Mathematics, LStat and LRisk, KU Leuven, Belgium, and Dept. of Mathematical Statistics and Actuarial Science, Univ. of the Free State, South Africa.\\
%		E-Mail: jan.beirlant@kuleuven.be\\}
%\end{flushleft}	
%	
%	\begin{flushleft}
%Zhengxiao Li\\
% \emph{School of Insurance and Economics,\\
%University of International Business and Economics\\
%Beijing 100029, China\\
%E-Mail: li\_zhengxiao@126.com\\}
%\end{flushleft}
%
%\begin{flushleft}
%Shengwang Meng (Corresponding author)\\
% \emph{Center for Applied Statistics and School of Statistics\\
%Renmin University of China\\
%Beijing 100872, China\\
%E-Mail: mengshw@ruc.edu.cn\\}
%\end{flushleft}

\bibliography{mybibfile}

\end{document}